\documentclass[PDF]{article} 
\usepackage{graphicx}
\usepackage{layout}
\usepackage{amsmath,amssymb}
\usepackage{textcomp}
\usepackage{hyperref}

\newcommand{\be}{\begin{equation}}
\newcommand{\ee}{\end{equation}}
\newcommand{\bea}{\begin{eqnarray}}
\newcommand{\eea}{\end{eqnarray}}
\newcommand{\bwt}{\begin{widetext}}
\newcommand{\ewt}{\end{widetext}}

\newcommand{\aeq}{&=&}

\newcommand{\itDelta}{{\it \Delta}}

\newcommand{\bra}{\langle}
\newcommand{\ket}{\rangle}
\newcommand{\dbra}{\bra \! \bra}
\newcommand{\dket}{\ket \! \ket}

\newcommand{\me}{\mbox{e}}

\newcommand{\bq}{{\bar q}}

\newcommand{\rS}{{\rm S}}

\begin{document}

\setlength{\baselineskip}{18pt}

\addtolength{\jot}{5pt}
\addtolength{\abovedisplayskip}{5pt}
\addtolength{\belowdisplayskip}{5pt}

\begin{titlepage}
\title{\huge Towards a separation of the elements in turbulence\\ 
via the analyses within MPDFT
\vspace{1cm}\\}


\author{{\Large Toshihico~Arimitsu$^{*1}$,  
        Naoko~Arimitsu$^{*2}$,     
        Kohei~Takechi$^{*1}$}\\ {\Large Yukio~Kaneda$^{*3}$ and       
        Takashi~Ishihara$^{*3}$}
\vspace*{1cm}\\
{$^{*1}$\ Faculty of Pure and Applied Sciences, University of Tsukuba}\\
		{305-8571 Tsukuba, Japan}\\
{$^{*2}$\ Graduate School of Environment and Information Sciences,
		Yokohama Nat'l University}\\ {240-8501 Yokohama, Japan}\\
{$^{*3}$\ Graduate School of Engineering, Nagoya University,}\\
     {464-8603 Nagoya, Japan}
}

\date{\empty}
\maketitle

\vspace{1cm}

\begin{abstract}
The PDFs for energy dissipation rates created in a high resolution 
from $4096^3$ DNS for fully developed turbulence are 
analyzed in a high precision with the PDF derived within the formula
of multifractal probability density function theory (MPDFT).
MPDFT is a statistical mechanical ensemble theory constructed in order to analyze 
intermittent phenomena through the experimental PDFs with fat-tail.
By making use of the obtained w-PDFs created from the whole of the DNS region, 
analyzed for the first time are the two partial PDFs, i.e.,
the max-PDF and the min-PDF which are, respectively, taken out from 
the partial DNS regions of the size $512^3$ with maximum and minimum enstropy.
The main information for the partial PDFs are the following. 
One can find a w-PDF whose tail part can adjust the slope of the tail-part of a max-PDF
with appropriate magnification factor.
The value of the point at which the w-PDF multiplied by the magnification factor starts
to overlap the tail part of the max-PDF coincides with the value of the connection point
for the theoretical w-PDF.
The center part of the min-PDFs can be adjusted quite accurately by the scaled w-PDFs
with a common scale factor.
\end{abstract}

\thispagestyle{empty}
\end{titlepage}


\section{Introduction
\label{Intro}}

There are the keystone works~\cite{Mandelbrot74,Frisch-Parisi83,Benzi-Paladin-Parisi-Vulpiani84,%
Halsey-Jensen-Kadanoff-Procaccia-Shraiman86,Meneveau87b,Nelkin90,Hosokawa91,%
Benzi-Biferale-Paladin-Vulpiani-Vergassola91,She-Leveque94,AA,AA1,AA4,AA5,AA6,%
Biferale_etal04,Chevillard06}
providing the multifractal aspects for fully developed turbulence.
Most of the works~\cite{Mandelbrot74,Frisch-Parisi83,Benzi-Paladin-Parisi-Vulpiani84,%
Halsey-Jensen-Kadanoff-Procaccia-Shraiman86,Meneveau87b,Nelkin90,Hosokawa91,%
She-Leveque94,AA,AA1}
deal with the scaling property of the system, e.g., comparison of
the scaling exponents of velocity structure function.
Only a few~\cite{Benzi-Biferale-Paladin-Vulpiani-Vergassola91,%
AA4,AA5,AA6,Biferale_etal04,Chevillard06
}
analyze the probability density functions (PDFs) for physical quantities
representing intermittent character.
Among these researches, multifractal probability density function theory 
(MPDFT)~\cite{AA4,AA5,AA6,AA17,AA18} 
provides the most precise analysis of fat-tail PDFs, 
which is a statistical mechanical ensemble theory constructed by the authors 
(T.A.\ and N.A.) in order to analyze intermittent phenomena through fat-tail PDFs 
extracted out from the data by experiments or numerical simulations.

In order to extract the intermittent character of 
the fully developed turbulence, it is necessary to have
information of hierarchical structure of the system,
which is realized by producing a series of PDFs for 
responsible singular quantities with different lengths 
\be
\ell_{n}= \ell_0 \delta^{-n}, \quad \delta>1
\quad ({n}=0,1,2,\cdots)
\label{def of delta}
\ee
that characterize the regions in which the physical quantities are coarse-grained.
The value for $\delta$ is chosen freely by observers.
NETFD is constructed such a way that the choice of $\delta$ should not affect 
the theoretical estimation of the values for the fundamental quantities 
characterizing the turbulent system under consideration.

Within MPDFT, it is assumed that there are two contributions to form 
a fat-tail PDF, i.e., one is a coherent contribution coming from a coherent 
turbulent motion and the other is a incoherent contribution from 
the dissipation term in the N-S equation which violates the invariance under
the scale transformation given at the beginning of the next section.
It is also assumed that the fat-tail PDF can be divided into two parts, i.e.,
the tail part and the center part.
As the tail part is responsible for the intermittent character of the system,
the coherent contribution may dominate over the incoherent contribution
at the tail part.
On the other hand, the center part consists of both coherent and incoherent contributions.
With the above consideration, we neglect within A\&A model of MPDFT 
the incoherent contribution to the tail part in deriving the theoretical PDF formula,
and conjecture that the coherent contribution to the tail part can be determined solely 
by the PDF for $\alpha$ provided in the next section.
Through the conjecture, we are able to separate theoretically the incoherent contribution 
from the coherent one at the center part.

In this paper, we analyze in a high accuracy the PDFs for energy dissipation rates
created in a high resolution from the snapshot raw data of $4096^3$ DNS 
for fully developed turbulence conducted by 
the authors (Y.K.\ and T.I.)~\cite{Kaneda-Ishihara05}. 
By making use of the result obtained by the precise analyses on the PDFs obtained from 
the whole of the DNS region, we analyze for the first time those PDFs extracted 
from the partial DNS regions of the size $512^3$ obtained by cutting 
the whole of $4096^3$ DNS region into $512$ pieces.
In Section~\ref{A&A model}, a brief introduction of A\&A model is given with its concept 
and basics.
In Section~\ref{PDFs A&A model within MPDFT}, the formula for the theoretical PDF 
for energy dissipation rates is derived within A\&A model of MPDFT.
The two distinct divisions of the PDF are introduced, which are important 
in the following analyses.
In Section~\ref{analysis of whole PDF}, the high-resolution PDFs created from 
the whole of $4096^3$ DNS region are analyzed in a high precision 
with the derived theoretical PDF.
In Section~\ref{dense and rare PDFs}, the partial PDFs created from 
the $512^3$ partial DNS regions with maximum enstrophy and minimum enstrophy
are studied with the help of the PDFs extracted from the whole of $4096^3$ DNS region.
Conclusion is provided in Section~\ref{conclusion}.

\section{A\&A model
\label{A&A model}}

It is known that the Navier-Stokes (N-S) equation 
$
\partial {\vec u}/\partial t
+ ( {\vec u}\cdot {\vec \nabla} ) {\vec u} 
= - {\vec \nabla} p
+ \nu \nabla^2 {\vec u}
$
for an incompressible fluid (${\vec \nabla} \cdot {\vec u} =0$)
is invariant under the scale transformation 
$
{\vec x} \rightarrow \lambda {\vec x}
$
accompanied by the rescaling in velocity field, time and pressure, i.e.,
$
{\vec u} \rightarrow \lambda^{\alpha/3} {\vec u}
$,
$
t \rightarrow \lambda^{1- \alpha/3} t
$
and
$
p \rightarrow \lambda^{2\alpha/3} p
$
with an arbitrary real number $\alpha$, when $\nu = 0$.
Here, $\nu$ is the kinematic viscosity;
${\vec u}$ is the velocity field; 
$p$ is the pressure of fluid per mass density.
It is assumed that for high Reynolds number ($\nu \rightarrow 0$) 
the singularities distribute themselves in a multifractal way in real physical space,
and that they are the origin of a coherent turbulent motion providing the fat-tail part of PDFs 
\cite{Frisch-Parisi83}.
In treating an actual turbulent system, the value of $\nu$ is fixed to 
a finite non-zero value unique to the material of fluid prepared for an experiment. 
Therefore, for the study of fully developed turbulence, we have to look for 
the characteristics of a coherent turbulent motion surrounded by an incoherent fluctuating motion.
The latter motion is due to the dissipation term $\nu \nabla^2 {\vec u}$ in the N-S equation 
which is the term violating the invariance under the scale transformation.
MPDFT is a statistical mechanical ensemble theory for analyzing both 
the coherent turbulent motion and the incoherent fluctuating motion.
The dissipation term can become effective depending on the region under consideration
since the spots of the region where the invariance is broken 
locate here and there in fully developed turbulence, and provide us with 
a specific character of fluctuation around the coherent motion.

Let us consider the energy dissipation rate $\varepsilon_n$ that is related to $\alpha$ by
\be
\varepsilon_n = \left(\ell_n / \ell_0 \right)^{\alpha-1}.
\label{x'-alpha transf}
\ee
Note that $\lim_{n \rightarrow \infty} \varepsilon_n \rightarrow \infty$ for
$
\alpha < 1
$.
The degree of singularity for energy dissipation rate is specified by 
the singularity exponent $\alpha$ which is considered to be a stochastic variable
whose PDF, $P^{(n)}(\alpha)$, is given by~\cite{AA,AA1,AA4,AA5,AA6,AA17,AA18}  
the R\'enyi~\cite{Renyi} or HCT~\cite{Havrda-Charvat,Tsallis88} type PDF
\be
P^{(n)}(\alpha) \propto \left[ 1  - (\alpha - \alpha_0)^2 / (\itDelta \alpha)^2
\right]^{n/(1-q)} 
\label{Tsallis prob density}
\ee
with
$
\itDelta \alpha = \sqrt{2X / (1-q) \ln \delta}
$.
This is the MaxEnt PDF of the R\'enyi entropy or the HCT entropy
\footnote{
The function (\ref{Tsallis prob density}) is the MaxEnt PDF derived from the R{\'e}nyi entropy
or from the HCT entropy with two constraints, one is the normalization condition
and the other is a fixed $q$-variance~\cite{Tsallis88}.
}.
The domain of $\alpha$ is
$\alpha_{\rm min} \leq \alpha \leq \alpha_{\rm max}$ with
$\alpha_{\rm min}$ and $\alpha_{\rm max}$ being given by
$
\alpha_{\rm min/ max} = \alpha_0 \mp \itDelta \alpha 
$.
$q$ is the entropy index.
The multifractal spectrum introduced through 
$
P^{(n)}(\alpha) \propto (\ell_n / \ell_0 )^{1-f(\alpha)}
$
is given for $n \gg 1$ by
\be
f(\alpha) = 1 + \left\{\ln \left[ 1 - 
(\alpha - \alpha_0)^2/(\itDelta \alpha )^2 \right]\right\} / (1-q)\ln \delta.
\label{Tsallis f d-alpha}
\ee

The three parameters $\alpha_0 = \bra \alpha \ket$, 
$X$ and $q$ are determined by the following three conditions
where the average $\bra \cdots \ket$ is taken with $P^{(n)}(\alpha)$.
One is the energy conservation law 
$
\bra \varepsilon_n \ket = \epsilon
$.
Another condition is the definition of the intermittency exponent $\mu$, i.e.,
$
\bra (\varepsilon_n/\epsilon )^2 \ket
= (\ell_n/\ell_0 )^{-\mu}
$.
The third condition is the new scaling relation
\be
\ln 2/(1-q)\ln \delta = 1/\alpha_- - 1/\alpha_+
\label{new scaling relation}
\ee
with $\alpha_\pm$ being the solution of $f(\alpha_\pm) =0$, i.e.,
$
\alpha_\pm = \alpha_0 \pm (2b X)^{1/2}
$
with
$
b = (1 - \me^{-(1-q)\ln \delta})/(1-q)\ln \delta
$.
The scaling relation (\ref{new scaling relation}) is solved to give
$
(2bX)^{1/2} = -(1-q) \log_2 \delta 
+ \{\alpha_0^2 + [(1-q) \log_2 \delta ]^2\}^{1/2}
$.
The new scaling relation is a generalization of the one 
introduced by Tsallis and others~\cite{Costa,Lyra98} 
to which (\ref{new scaling relation}) reduces when $\delta=2$.
This generalization was born out of A\&A model 
within the theoretical framework of MPDFT itself. 
Note that the A\&A model is a one-parameter model depending only on $\mu$.

The parameter $q$ is determined, altogether with $\alpha_0$ and $X$, 
as a function of $\mu$ only in the combination $(1-q)\ln \delta$.
It is quite reasonable in the following reason.
The value of the magnification $\delta$ is determined arbitrarily by observers, 
therefore its value should not affect the values of physical quantities as long as
one studies a single turbulent system.
The difference in $\delta$ is absorbed into the entropy index $q$,
therefore changing the zooming rate $\delta$ 
may result in picking up the different hierarchy, containing the entropy
specified by the index $q$, out of multifractal structure of turbulence.

\section{PDFs of energy dissipation rates with A\&A model
\label{PDFs A&A model within MPDFT}}

Let us create the probability $\Pi^{(n)}(\varepsilon_n) d\varepsilon_n$ 
to find the physical quantity 
$
\varepsilon_n
$
taking a value in the domain $\varepsilon_n \sim \varepsilon_n+d\varepsilon_n$,
whose normalization is specified by
$
\int_{0}^{\infty} d\varepsilon_n  \Pi^{(n)}(\varepsilon_n) =1
$.
We assume that the PDF, $\Pi^{(n)}(\varepsilon_n)$,
can be divided into two parts as
\be
\Pi^{(n)}(\varepsilon_n) = \Pi^{(n)}_{\rS}(\varepsilon_n) 
+ \Delta \Pi^{(n)}(\varepsilon_n).
\label{def of Pi phi}
\ee
The first term is the part representing a coherent turbulent motion described in
the limit $\nu \rightarrow 0$.
The coherent contribution is assumed to be given by~\cite{AA4}
\be
\Pi^{(n)}_{\rS}(\vert \varepsilon_n \vert) d\vert \varepsilon_n \vert 
= \bar{\Pi}^{(n)}_{\rS} P^{(n)}(\alpha) d \alpha
\label{singular portion}
\ee
with the variable translation (\ref{x'-alpha transf}) 
between $\varepsilon_n$ and $\alpha$.
On the other hand, the second term $\Delta \Pi^{(n)}(\varepsilon_n)$ 
in (\ref{def of Pi phi})
represents an incoherent contribution from the dissipation term in the N-S equation.
The dissipation term violates the invariance under the scale transformation,
and therefore the incoherent contribution has not been taken into account 
in the consideration given in the previous section for $P^{(n)}(\alpha)$, i.e.,
the effect of the finiteness of the dissipation term is not included in 
$\Pi^{(n)}_{\rS}(\varepsilon_n)$.
Each term is composed of the product of two PDFs, i.e.,
1) the PDF that determines from which the contribution is originated out of 
two independent origins, i.e., the coherent origin or the incoherent origin, and
2) the conditional PDF to find a value $\varepsilon_n$ in the domain 
$\varepsilon_n \sim \varepsilon_n+d\varepsilon_n$ for each origin.

We divide the PDF $\Pi^{(n)}(\varepsilon_n)$ into two parts another way, i.e.,
\be
\Pi^{(n)}(\varepsilon_n) = \Pi^{(n)}_{\rm ct}(\varepsilon_n) 
+ \Pi^{(n)}_{\rm tl}(\varepsilon_n).
\label{def of Pi phi ct tl}
\ee
The tail part $\Pi^{(n)}_{\rm tl}(\varepsilon_n)$ for 
$\vert \varepsilon_n \vert \geq \varepsilon^*_n$ (equivalently, 
$\alpha_{\rm min} \leq \alpha \leq \alpha^*$) represents mainly the contribution 
of the singularities due to the coherent turbulent motion, 
whereas the center part $\Pi_{\phi,{\rm cr}}^{\prime (n)}(\varepsilon_n)$ for 
$\vert \varepsilon_n \vert \leq \varepsilon^*_n$ (equivalently, 
$\alpha_{\rm min} \geq \alpha \leq \alpha^*$) does 
the contributions from both coherent and incoherent motions.
Here, $\varepsilon^*_n$ is the connection point of the two PDFs,
$\Pi^{(n)}_{\rm ct}(\varepsilon_n)$ and 
$\Pi^{(n)}_{\rm tl}(\varepsilon_n)$, which is related to $\alpha^*$
through (\ref{x'-alpha transf}).
It is reasonable to assume that the origin of intermittent 
rare events is attributed to the first singular term 
in (\ref{def of Pi phi}), and that the contribution from 
the second term $\itDelta \Pi^{(n)}(\varepsilon_n)$ to the events is negligible.
We assume therefore that, for the tail part PDF
$\Pi^{(n)}_{\rm tl}(\varepsilon_n)$,
one can neglect completely the contribution from the second correction term in 
(\ref{def of Pi phi}).
Under this assumption, we put
\be
\Pi^{(n)}_{\rm tl}(\varepsilon_n) d\varepsilon_n 
= \bar{\Pi}^{(n)}_{\rS} P^{(n)}(\alpha) d \alpha
\label{Pi_tl P_S}
\ee
for $\varepsilon_n \geq \varepsilon^*_n$.

Let us introduce, here, the PDF $\hat{\Pi}^{(n)}(\xi_n)$, through the relation
$
\hat{\Pi}^{(n)}(\xi_n) d\xi_n = \Pi^{(n)}(\varepsilon_n) d \varepsilon_n
$,
for the new variable 
$
\xi_n = \varepsilon_n / \dbra \varepsilon_n^2 \dket_c^{1/2}
$
normalized by the standard deviation.
Here, $\dbra (\varepsilon_n)^2 \dket_c = \dbra (\varepsilon_n)^2 \dket - \dbra \varepsilon_n \dket^2$
and the average $\dbra \cdots \dket$ is taken with $\Pi^{(n)}(\varepsilon_n)$.
The normalized variable $\xi_n$ is related to $\alpha$ by
$
\xi_n = \bar{\xi}_n (\ell_n / \ell_0 )^{\alpha - \zeta_{6}/2}
$
with
$
\zeta_{3m} = 1-\tau(m)
$
and
$
\bar{\xi}_n = [\{ \gamma_{2}^{(n)} - [ \gamma_{1}^{(n)} + ( 1- \gamma_{0}^{(n)} )
a_3 (\ell_n / \ell_0 )^{\zeta_{3}} ]^2 \} 
(\ell_n / \ell_0 )^{-\zeta_{6}} + (1- \gamma_{0}^{(n)} ) a_{6} ]^{-1/2}
$
where
$
a_{3m} = (\vert f''(\alpha_0) / f''(\alpha_{m}) \vert )^{1/2}
$,
$
\gamma^{(n)}_{m} = (\ell_n/\ell_0)^m \int_{0}^{\infty} d\varepsilon_n 
( \varepsilon_n )^m \Delta \Pi^{(n)}(\varepsilon_n)
$
and  
$
\tau(\bq) = 1-\alpha_0 \bq + 2X\bq^2 /(1 + C_{\bq}^{1/2} )
+ [ 1-\log_2 (1+ C_{\bq}^{1/2} ) ] (\ln 2 ) / (1-q)\ln \delta 
$
with
$
{C}_{\bq}= 1 + 2 \bq^2 X (1-q) \ln \delta
$.
The mass exponent $\tau(\bq)$ is related to $f(\alpha)$ by
the Legendre transformation~\cite{Halsey-Jensen-Kadanoff-Procaccia-Shraiman86,Meneveau87b}
$
f(\alpha) = \alpha \bq + \tau(\bq)
$ with
$ 
\bq = d f(\alpha) / d \alpha
$ and
$
\alpha = - d \tau(\bq) / d \bq
$.
The variables $\bq$ and $\alpha$ are related through
$
\bq = [2/(1-q) \ln \delta] [(\alpha -\alpha_0)/ 
( \alpha -\alpha_{\rm max}) ( \alpha -\alpha_{\rm min})]
$.
The $\bq$th moment of the energy dissipation rate is given by
$
\bra (\varepsilon_n / \epsilon )^{\bq} \ket
\propto (\ell_n / \ell_0 )^{-\tau(\bq) + 1 - \bq}
$~\cite{Meneveau87b,AA17,AA18}.
With the assumption (\ref{Pi_tl P_S}), we have, for 
$\xi_n^* \leq \vert \xi_n \vert  \leq \xi_n^{\rm max}$ 
(equivalently, $\alpha_{\rm min} \leq \alpha \leq \alpha^*$) 
with
$
\xi_n^{\rm max} = \bar{\xi}_n (\ell_n/\ell_0)^{\alpha_{\rm min} - \zeta_{6} /2}
$,
\bea
\hat{\Pi}^{(n)}_{\rm tl}(\xi_n)
= \bar{\Pi}^{(n)}\ 
\left(\ell_n/\ell_0 \right)^{1 -f(\alpha)} \bar{\xi}_n /\xi_n
\label{PDF tail part}
\eea
where
$
\bar{\Pi}^{(n)} = \bar{\Pi}_{\rS}^{(n)} \bar{\xi}_n 
(\vert f^{\prime \prime}(\alpha_0) \vert/2\pi \vert \ln (\ell_n/\ell_0) \vert)^{1/2}
$
with
$
\bar{\Pi}^{(n)}_{\rS} = 1- \gamma_{0}^{(n)}
$.
Here, $\xi^*_n$ and $\alpha^*$ are related with each other by
$
\xi_n^* = \bar{\xi}_n (\ell_n/\ell_0)^{\alpha^* - \zeta_{6} /2}
$.

It may be appropriate to perform the connection by means of
$\hat{\Pi}^{(n)}_{\rm ct}(\xi_n)$ and $\hat{\Pi}^{(n)}_{\rm tl}(\xi_n)$.
The tail and the center parts of PDFs are connected at 
$
\xi_n^*
$
under the conditions that they have the common value, i.e.,
$
\hat{\Pi}^{(n)}_{\rm tl}(\xi_n^*)
= \hat{\Pi}^{(n)}_{\rm cr}(\xi_n^*)
$
and the common log-slope, i.e.,
$
d \ln \hat{\Pi}^{(n)}_{\rm tl}(\xi_n) /d\xi_n
\vert_{\xi_n=\xi_n^*}
= d \ln \hat{\Pi}^{(n)}_{\rm cr}(\xi_n) /d\xi_n
\vert_{\xi_n=\xi_n^*}.
$
The value $\xi_n^*$ (therefore,  $\varepsilon^*_n$) 
is determined for each PDF as an adjusting parameter in the analysis of PDFs,
and provides us with an important information of the border line
beyond which the contribution from the incoherent motion ceases.

As there is no theory at present to produce the formula of $\hat{\Pi}^{(n)}_{\rm cr}(\xi_n)$ 
for $\vert \xi_n \vert \leq \xi_n^*$ (equivalently, $\alpha^* \leq \alpha$),
we introduce a trial PDF which, after the connection procedure, is given by
\bea
\hat{\Pi}^{(n)}_{\rm cr}(\xi_n) 
\aeq \bar{\Pi}^{(n)} \ (\ell_n/\ell_0)^{1 -f(\alpha^*)}
\ \me^{-[g(\xi_n) - g(\xi_n^*)]}\ \bar{\xi}_n/\xi_n^*
\label{PDF center part}
\eea
with a trial function of the Tsallis-type
$
\me^{-g(\xi_n)} = \left(\xi_n / \xi_n^* \right)^{\theta - 1}
\left\{1- \left(1-q^\prime \right)
\left[\theta+f^\prime(\alpha^*) \right]
\left[\left(\xi_n / \xi_n^* \right)^{w} -1 \right] / w \right\}^{1/(1-q^\prime)}
$
and
$
\bar{\Pi}^{(n)} = \bar{\Pi}_{\rS}^{(n)} 
\sqrt{\vert f^{\prime \prime}(\alpha_0) \vert / 2\pi \vert \ln (\ell_n/\ell_0) \vert} / \bar{\xi}_n
$. 
The parameter $w$ is adjusted by the property of the experimental PDFs 
around the peak point; 
$q^\prime$ is the entropy index different from $q$ 
in (\ref{Tsallis prob density});
$\theta$ is determined by the property of PDF near $\xi_n = 0$.
The contribution to $\hat{\Pi}^{(n)}_{\rm cr}(\xi_n)$ comes
both from coherent and incoherent motions.

Thanks to the assumption that, for the tail part
$\xi_n^* \leq \xi_n \leq \xi_n^{\rm max}$, 
one can neglect the contribution from 
$\Delta \Pi^{(n)}(x_n)$, 
we obtain the formula to calculate $\gamma_{m}^{(n)}$ in the expression
$
\gamma_{m}^{(n)} 
= (K_{m}^{(n)} - L_{m}^{(n)})/
(1+K_{0}^{(n)} - L_{0}^{(n)})
$
with
\bea
K_{m}^{(n)} \aeq  (\ell_n/\ell_0)^{1-f(\alpha^*) + m \alpha^*}
\sqrt{\vert f^{\prime \prime}(\alpha_0) \vert/2\pi \vert \ln \left(\ell_n/\ell_0 \right) \vert}
\int_0^1 dz\ z^m 
\me^{-[g(\xi_n^* z)-g(\xi_n^*)]},
\label{K}
\\
L_{m}^{(n)} \aeq (\ell_n/\ell_0) 
\sqrt{\vert f^{\prime \prime}(\alpha_0) \vert
\vert \ln \left(\ell_n/\ell_0\right) \vert/2\pi} 
\int_{\alpha^*}^{\alpha_{\rm max}} d\alpha \ 
(\ell_n/\ell_0)^{m \alpha - f(\alpha)}.
\label{L}
\eea

\section{Analysis of PDFs taken from the whole of $4096^3$ DNS region
\label{analysis of whole PDF}}

PDFs of energy dissipation rates studying in this section are extracted by cooking 
the snapshot data taken from the whole of $4096^3$ DNS region.
The DNS~\cite{Kaneda-Ishihara05} is specified by the Taylor micro-scale Reynolds number 
$R_\lambda = 1132$ and the integral length $L/\eta = 2.130 \times 10^3$ with 
the Kolmogorov length scale 
$
\eta = 5.12 \times 10^{-4}
$.
The inertial range is estimated to be in the region $126 \lesssim r/\eta \lesssim 448$.

\begin{figure}
\begin{center}
\vspace*{-2.5cm}
\resizebox*{8.0cm}{!}{\includegraphics{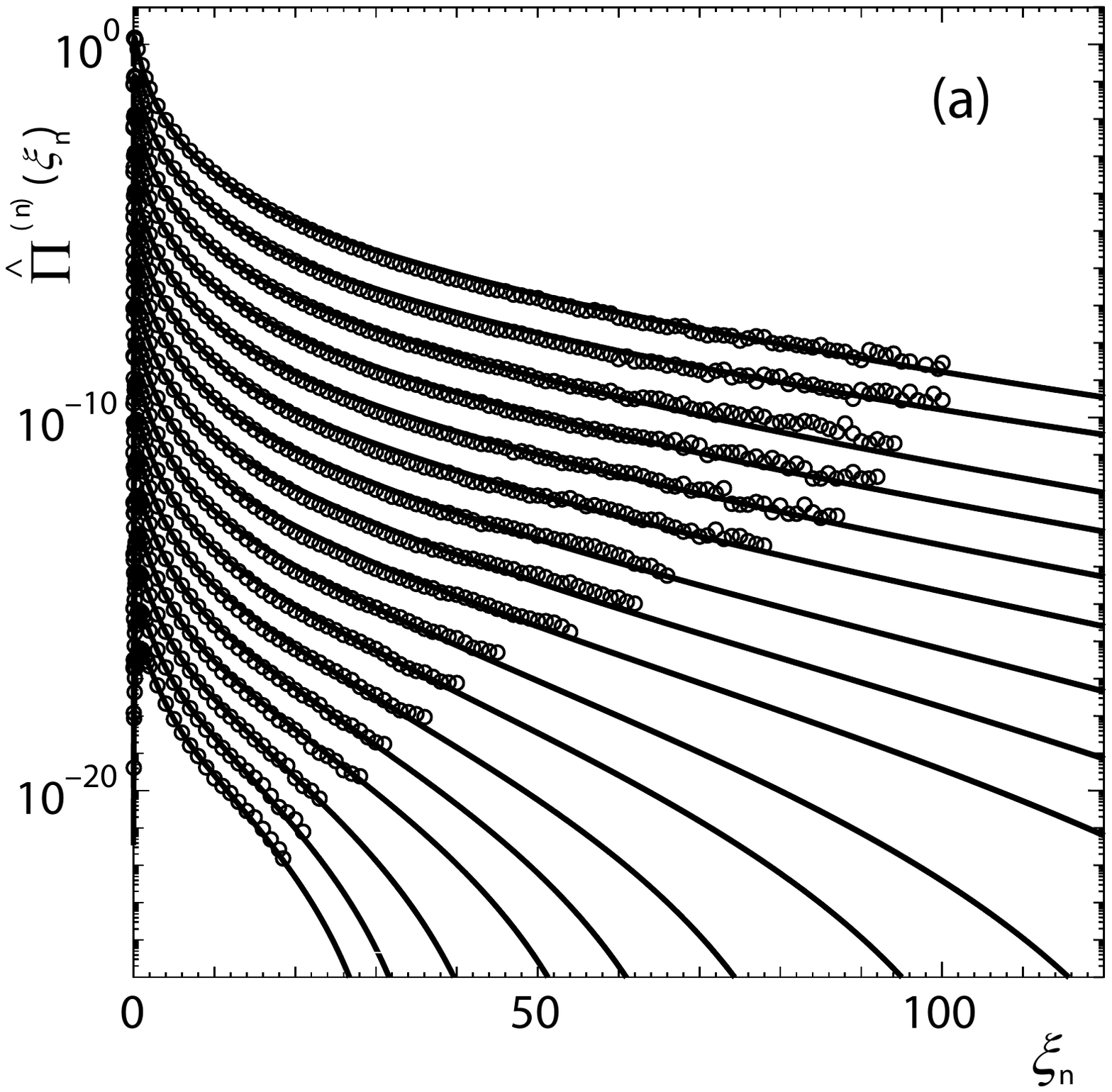}}%
\hspace*{0.5cm}
\resizebox*{8.0cm}{!}{\includegraphics{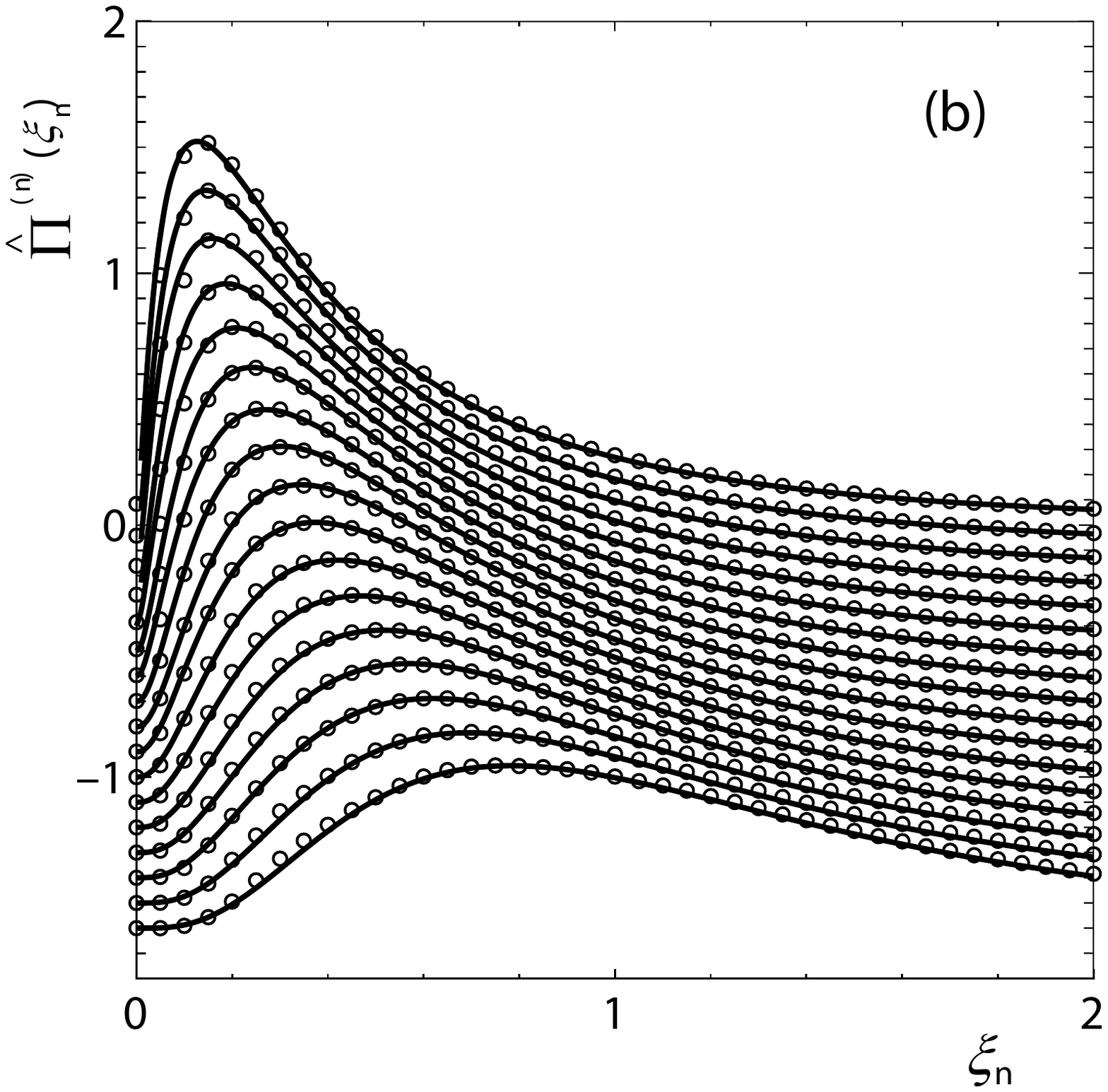}}%
\vspace{-4.8cm}
\hspace*{-0.5cm}
\resizebox*{8.0cm}{!}{\includegraphics{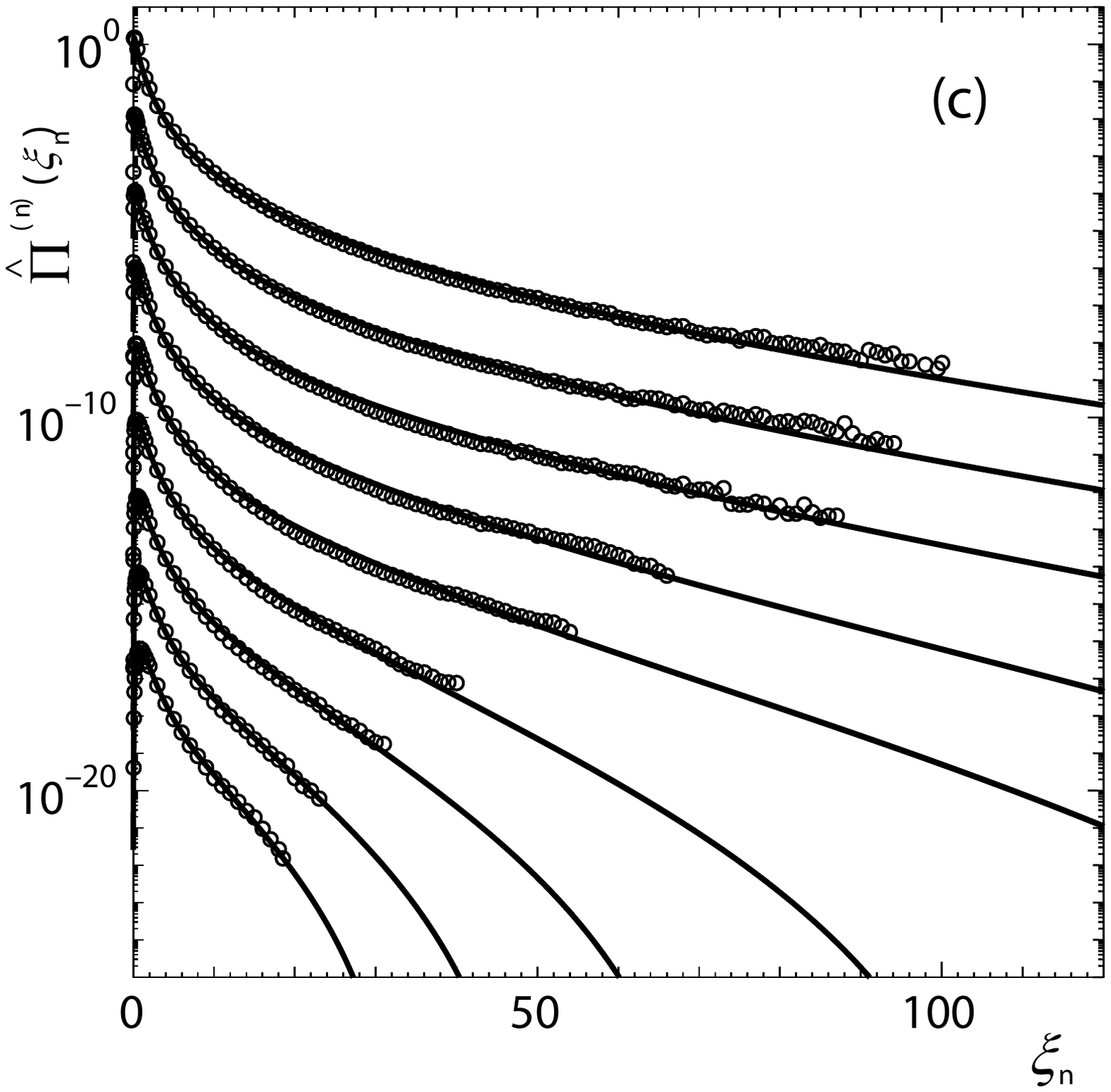}}%
\hspace*{-0.45cm}
\resizebox*{8.0cm}{!}{\includegraphics{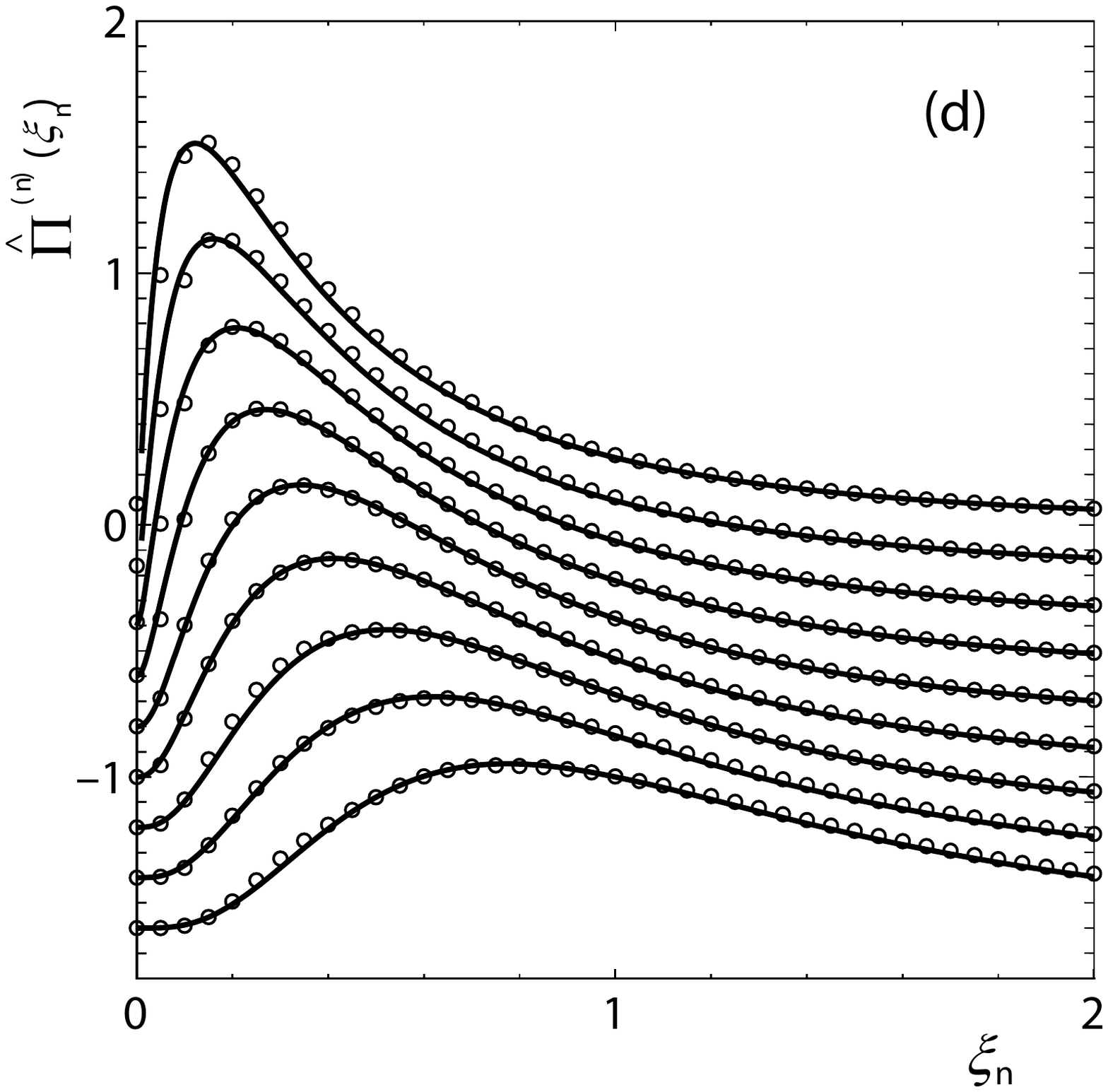}}%
\vspace{-4.8cm}
\hspace*{-0.5cm}
\resizebox*{8.0cm}{!}{\includegraphics{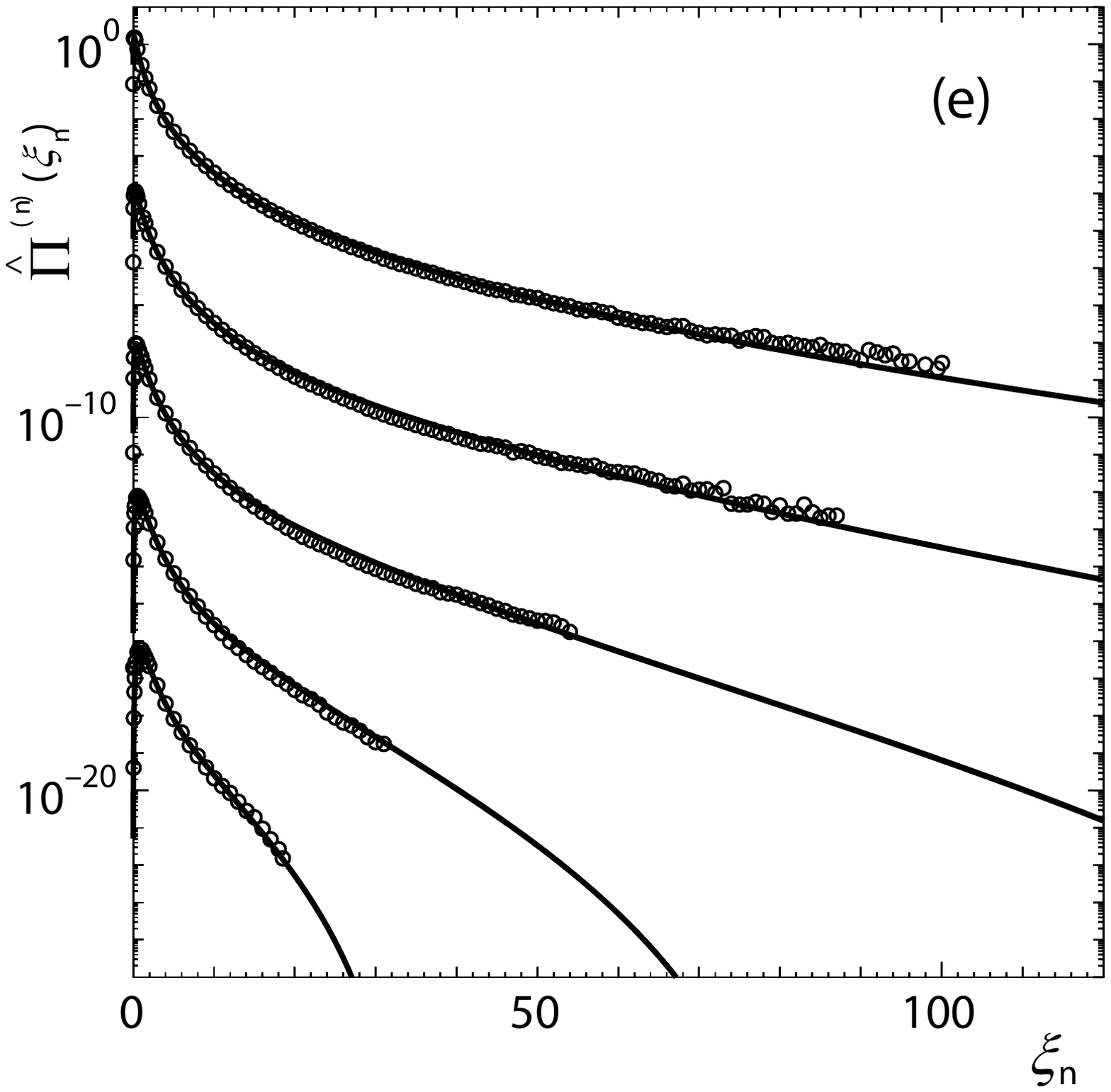}}%
\hspace*{-0.4cm}
\resizebox*{8.0cm}{!}{\includegraphics{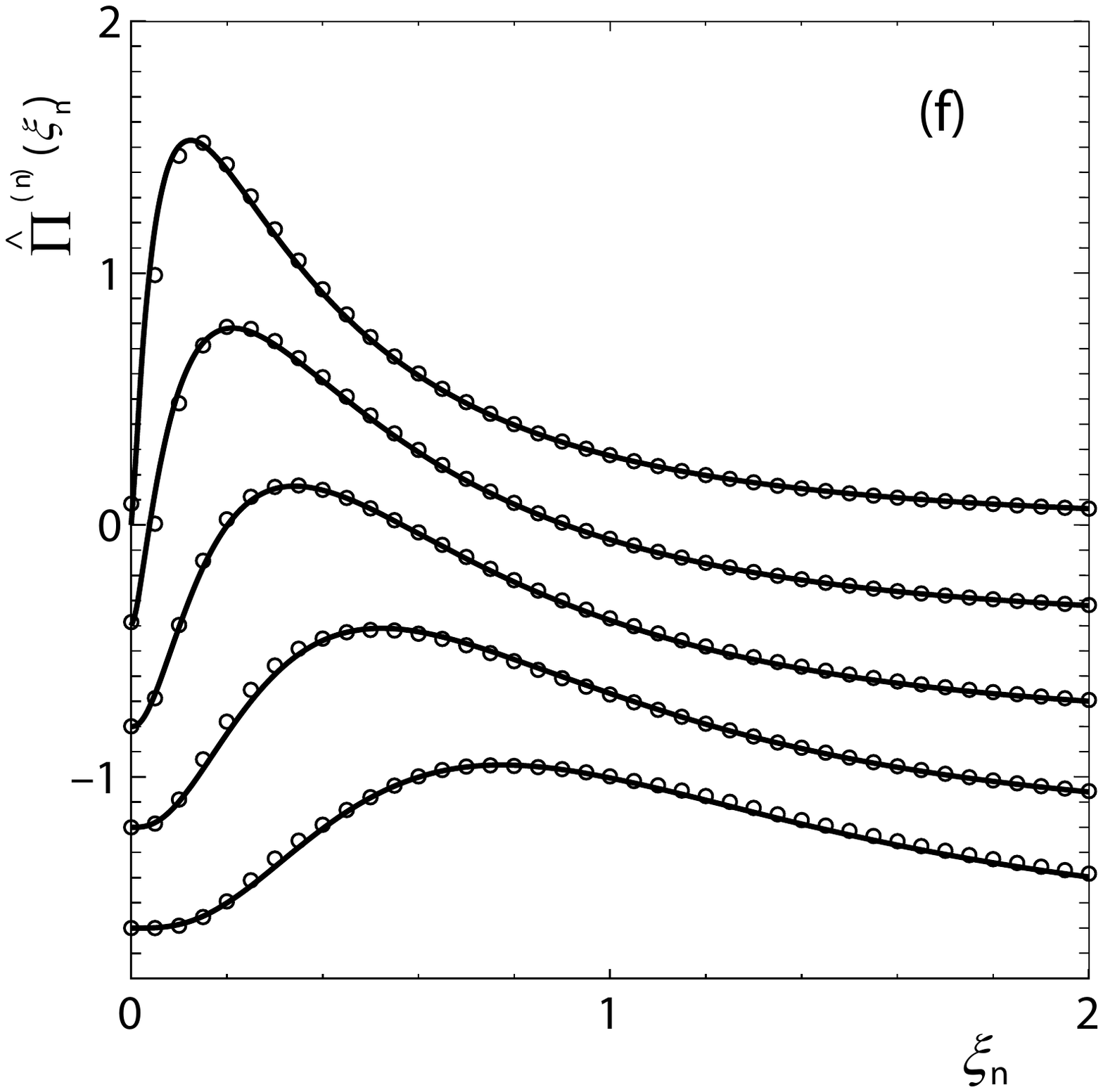}}%
\vspace{-2.8cm}
\caption{PDFs of energy dissipation rates (a), (b) for $\delta = 2^{1/4}$, 
(c), (d) for $\delta = 2^{1/2}$ and (e), (f) for $\delta = 2$.
Open circles represent the PDFs extracted out from the whole of $4096^3$ DNS region.
For better visibility, the PDFs in each figure are shifted along the vertical axes 
by the amount $-1$ in (a), $-0.1$ in (b), $-2$ in (c), $-0.2$ in (d), $-4$ in (e) and $-0.4$ in (f).
They are displayed according to the values of $r/\eta$ ($=\ell_n/\eta$) successively 
(see Table~\ref{pdf edr parameters1}) in the sequence that the PDF with the smallest value 
of $r/\eta$ is placed at the top and the one with the largest value at the bottom 
for each $\delta$. 
Solid lines represent the curves given by theoretical PDFs of A\&A model.
It turns out that $\mu = 0.345$ ($(1-q)\ln \delta = 0.299$, $\alpha_0=1.20$, $X=0.411$)
for every $\delta$.
Other parameters are listed in Table~\ref{pdf edr parameters1}.
The vertical axes in (a), (c), (e) are of log-scale, whereas those in (b), (d), (f) are 
of linear-scale.
\label{fig:pdf edr delta_2^1/4,^1/2,^1}}
\end{center}
\end{figure}

The PDFs are analyzed 
in Fig.~\ref{fig:pdf edr delta_2^1/4,^1/2,^1} for three different magnifications 
(a), (b) $\delta = 2^{1/4}$, (c), (d) $\delta = 2^{1/2}$ and (e), (f) $\delta = 2$.
The vertical axes of (a), (c), (e) in Fig.~\ref{fig:pdf edr delta_2^1/4,^1/2,^1}
are given in log scale which are good to see the tail parts of PDFs,
whereas those of (b), (d), (f) in the figure are given in linear scale
which are appropriate to observe the center parts of PDFs.
For better visibility, 
PDFs in each figure are shifted, successively, with respect to the values of $r/\eta$
by appropriate amounts along the vertical axis, i.e.,
$-1$ in (a), $-0.1$ in (b), $-2$ in (c), $-0.2$ in (d), $-4$ in (e) and $-0.4$ in (f).
The PDFs with smaller values of $r/\eta$ are placed at upper parts in each figure
and therefore PDFs with larger values of $r/\eta$ at lower parts. 
Open circles in the figures are the PDFs constructed with the help of 
the fluid velocity data taken from all the points in the whole of $4096^3$ DNS region.
Solid lines represent the theoretical PDFs.

In \cite{AA18}, the authors (N.A.\ and T.A.) performed, by means of the theoretical PDF,
the analysis of the PDFs furnished by Kaneda and Ishihara \cite{Kaneda-Ishihara05}
which had been created by setting the width of bins to be $0.2$ along the $\xi_n$ axis.
The resolution of the PDFs is good enough to analyze the tail part but is not enough
to analyze the center part (see Fig.~1, Fig.~2 and Fig.~3 in \cite{AA18}), which 
results in the difficulty in drawing precise theoretical curves for the center-part PDFs.
This ambiguity causes the scattering of parameters, especially, 
in the $r/\eta$-dependence of $\theta$ (see Fig.~7 in \cite{AA18}).
Note that $r$ ($= \ell_n$) in the present paper corresponds to $2r$ ($= \ell_n$)
in \cite{Kaneda-Ishihara05,AA18}.

In order to raise the resolution of PDFs, in the present paper, we created PDFs
by cooking the row data of $4096^3$ DNS turbulence.
In creating the tail-part PDFs in Fig.~\ref{fig:pdf edr delta_2^1/4,^1/2,^1},
we set the width of bins to be $5 \times 10^{-2}$ along the $\xi_n$ axis,
while, in creating the center-part PDFs, we set the width of bins to be $5 \times 10^{-3}$. 
Note that in drawing the tail-part (center-part) PDFs, not all the bins but
every $20$ ($10$) bins are plotted for better visibility.
We discarded the bins containing the number of data points less than 
$10^{-2}$ \% of the mean number of data points per bin on average.
For example, the bins containing less than $1.37 \times 10^2$ ($9.73 \times 10^4$) 
data points are discarded for $r/\eta = 27.5$ ($r/\eta = 443$). 
Note that there are $4096^3 = 6.87 \times 10^{10}$ data points as a whole.

The parameters necessary for the theoretical PDF within A\&A model, i.e.,
those for the tail part of the PDF (\ref{PDF tail part}) 
and for the center part of the PDF (\ref{PDF center part}),
are obtained through the analysis of the high resolution PDFs created from $4096^3$ DNS.
It turns out that the tail-part PDFs for the turbulent system under consideration
are characterized with the value of the intermittency exponent $\mu = 0.345$.
Then, the parameters necessary for the PDFs are determined as 
$(1-q)\ln \delta = 0.299$, $\alpha_0=1.20$ and $X=0.411$,
which are independent of $\delta$.
It may be worthwhile to note here that the entropy index becomes 
$q=-0.728$ for $\delta= 2^{1/4}$ ($= 1.19$),
$q=0.136$ for $\delta = 2^{1/2}$ ($= 1.41$) and $q=0.568$ for $\delta = 2$.
The parameters necessary for the center-part PDFs are
listed in Table~\ref{pdf edr parameters1} for each $\delta$.
Note that the value of $\mu$ extracted from the PDFs with high resolution in the present paper
and those with low resolution in  \cite{AA18} turns out to be the same, and there is 
no difference observed between the tail-part PDFs of two resolutions.

\begin{table}
\caption{The values of parameters in the theoretical PDF which are
obtained in the course of the analyses of PDFs extracted out from 
the whole of $4096^3$ DNS region for the cases
$\delta = 2^{1/4}$, $\delta = 2^{1/2}$ and $\delta = 2$.
\label{pdf edr parameters1}} 
\begin{center}
\scalebox{0.65}{
\begin{tabular}{c||c|c|c|c|c|c|c||c|c|c|c|c|c|c||c|c|c|c|c|c|c}
& \multicolumn{7}{c||}{$\delta = 2^{1/4}$} & \multicolumn{7}{c||}{$\delta =2^{1/2}$}
& \multicolumn{7}{c}{$\delta=2$} \\ \hline \hline
$r/\eta$ & $n$ & $\tilde{n}$ & $q'$ & $w$ & $\theta$ & $\alpha^{*}$ & $\varepsilon_n^*$ & $n$ & $\tilde{n}$ & $q'$ & $w$ & $\theta$ & $\alpha^{*}$ & $\varepsilon_n^*$ & $n$ & $\tilde{n}$ & $q'$ & $w$ & $\theta$ & $\alpha^{*}$ & $\varepsilon_n^*$ \\ 
  \hline  \hline
27.5 & 29.0 & 5.03 & 1.06 & 0.480 & 2.90 & 0.220 & 2.84 & 14.0 & 4.85 & 1.05 & 0.450 & 2.90 & 0.210 & 2.63 & 8.00 & 5.55 & 1.05 & 0.450 & 2.90 & 0.220 & 2.28 \\ 
32.8 & 28.0 & 4.85 & 1.06 & 0.490 & 3.10 & 0.210 & 2.83 & - & - & - & - & - & - & - & - & - & - & - & - & - & - \\ 
38.9 & 26.5 & 4.59 & 1.06 & 0.500 & 3.00 & 0.200 & 2.39 & 13.5 & 4.68 & 1.06 & 0.510 & 3.00 & 0.210 & 2.28 & - & - & - & - & - & - & - \\  
46.3 & 26.0 & 4.51 & 1.06 & 0.320 & 3.30 & 0.200 & 2.32 & - & - & - & - & - & - & - & - & - & - & - & - & - & - \\  
55.1 & 25.0 & 4.33 & 1.08 & 0.600 & 3.00 & 0.200 & 2.16 & 12.5 & 4.33 & 1.08 & 0.600 & 3.00 & 0.200 & 2.17 & 6.30 & 4.37 & 1.07 & 0.600 & 3.00 & 0.200 & 2.09 \\  
65.6 & 24.0 & 4.16 & 1.07 & 0.600 & 3.40 & 0.200 & 1.97 & - & - & - & - & - & - & - & - & - & - & - & - & - & - \\ 
78.0 & 23.0 & 3.99 & 1.08 & 0.650 & 3.30 & 0.200 & 1.68 & 11.5 & 3.99 & 1.08 & 0.650 & 3.30 & 0.200 & 1.68 & - & - & - & - & - & - & - \\  
92.7 & 22.0 & 3.81 & 1.08 & 0.650 & 3.70 & 0.200 & 1.49 & - & - & - & - & - & - & - & - & - & - & - & - & - & - \\ 
110 & 21.0 & 3.64 & 1.08 & 0.700 & 3.80 & 0.200 & 1.34 & 10.7 & 3.71 & 1.08 & 0.700 & 3.80 & 0.200 & 1.34 & 5.30 & 3.67 & 1.08 & 0.720 & 3.80 & 0.200 & 1.38 \\ 
131 & 20.0 & 3.47 & 1.08 & 0.730 & 3.80 & 0.200 & 1.09 & - & - & - & - & - & - & - & - & - & - & - & - & - & - \\ 
156 & 19.0 & 3.29 & 1.09 & 0.770 & 3.85 & 0.190 & 0.997 &  9.5 & 3.29 & 1.08 & 0.750 & 3.90 & 0.200 & 0.920 & - & - & - & - & - & - & - \\  
186 & 18.0 & 3.12 & 1.09 & 0.800 & 3.90 & 0.200 & 0.808 & - & - & - & - & - & - & - & - & - & - & - & - & - & - \\  
221 & 17.0 & 2.95 & 1.09 & 0.850 & 4.10 & 0.190 & 0.733 &  8.50 & 2.95 & 1.09 & 0.850 & 4.30 & 0.200 & 0.700 & 4.30 & 2.98 & 1.08 & 0.780 & 4.60 & 0.180 & 0.753 \\  
264 & 16.0 & 2.77 & 1.10 & 0.900 & 4.20 & 0.180 & 0.681 & - & - & - & - & - & - & - & - & - & - & - & - & - & - \\ 
314 & 15.0 & 2.60 & 1.10 & 0.920 & 4.50 & 0.180 & 0.559 &  7.50 & 2.60 & 1.09 & 0.900 & 4.50 & 0.180 & 0.570 & - & - & - & - & - & - & - \\  
374 & 14.0 & 2.43 & 1.09 & 0.950 & 4.70 & 0.180 & 0.473 & - & - & - & - & - & - & - & - & - & - & - & - & - & - \\  
442 & 13.0 & 2.25 & 1.10 & 1.02 & 4.85 & 0.160 & 0.448 &  6.50 & 2.25 & 1.10 & 1.00 & 5.00 & 0.160 & 0.458 & 3.30 & 2.29 & 1.10 & 1.05 & 4.70 & 0.170 & 0.438 \\  
\end{tabular}
}
\end{center}
\end{table}

\begin{figure}
  \begin{center}
\vspace*{-2.2cm}
\resizebox*{7.0cm}{!}{\includegraphics{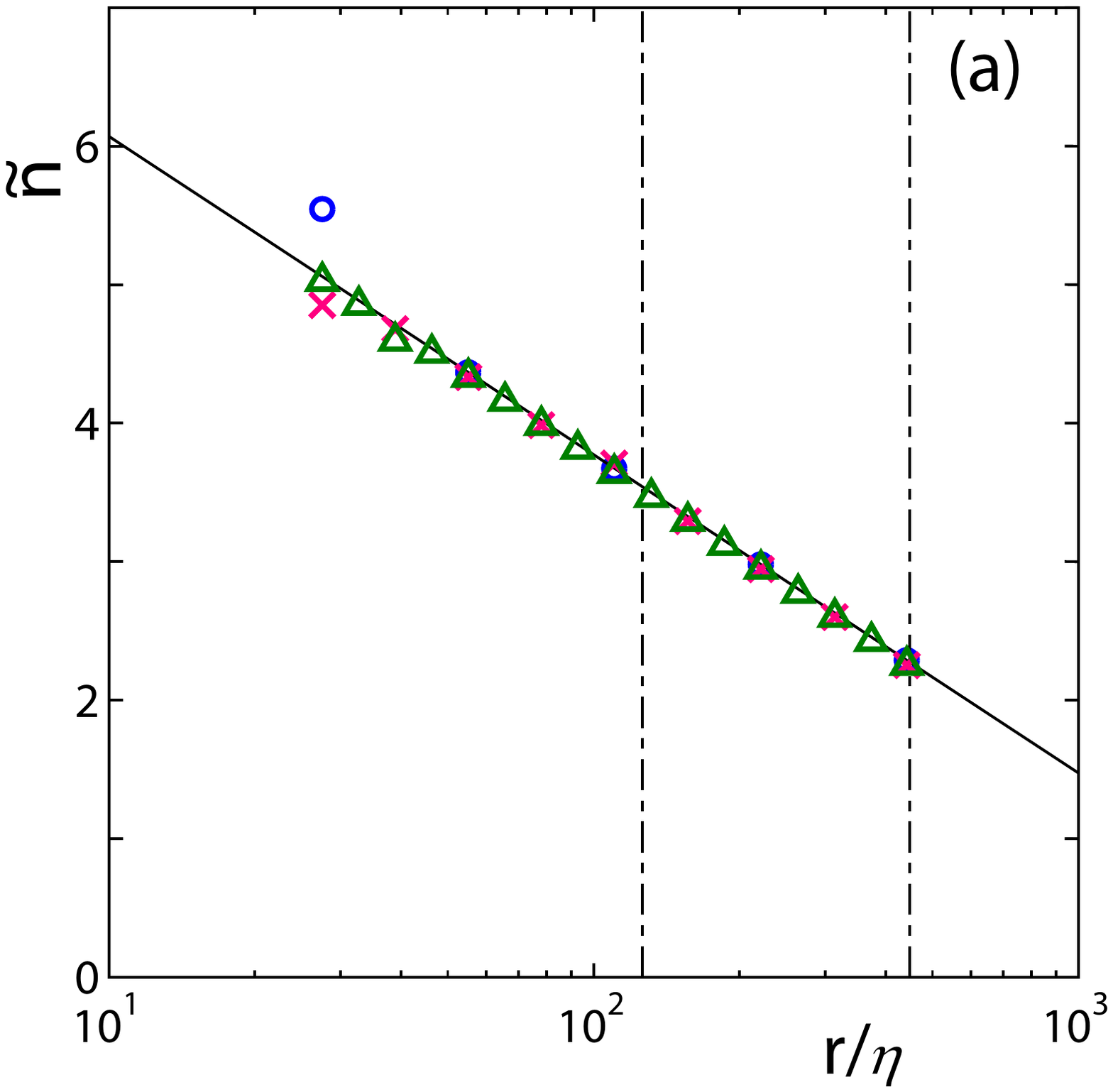}}%
\hspace*{0.5cm}
\resizebox*{7.0cm}{!}{\includegraphics{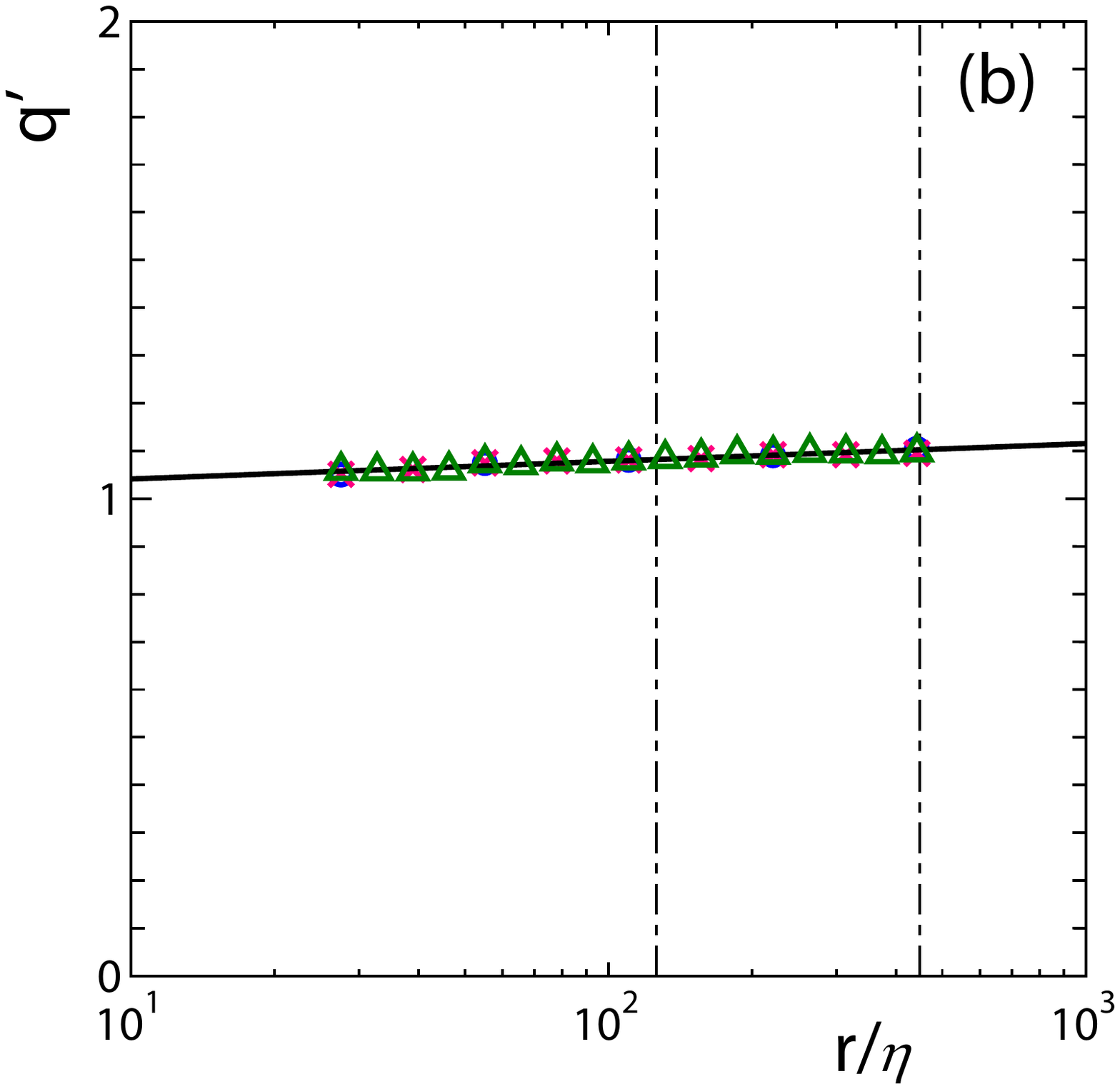}}%
\vspace{-4.0cm}
\hspace*{0.1cm}
\resizebox*{7.0cm}{!}{\includegraphics{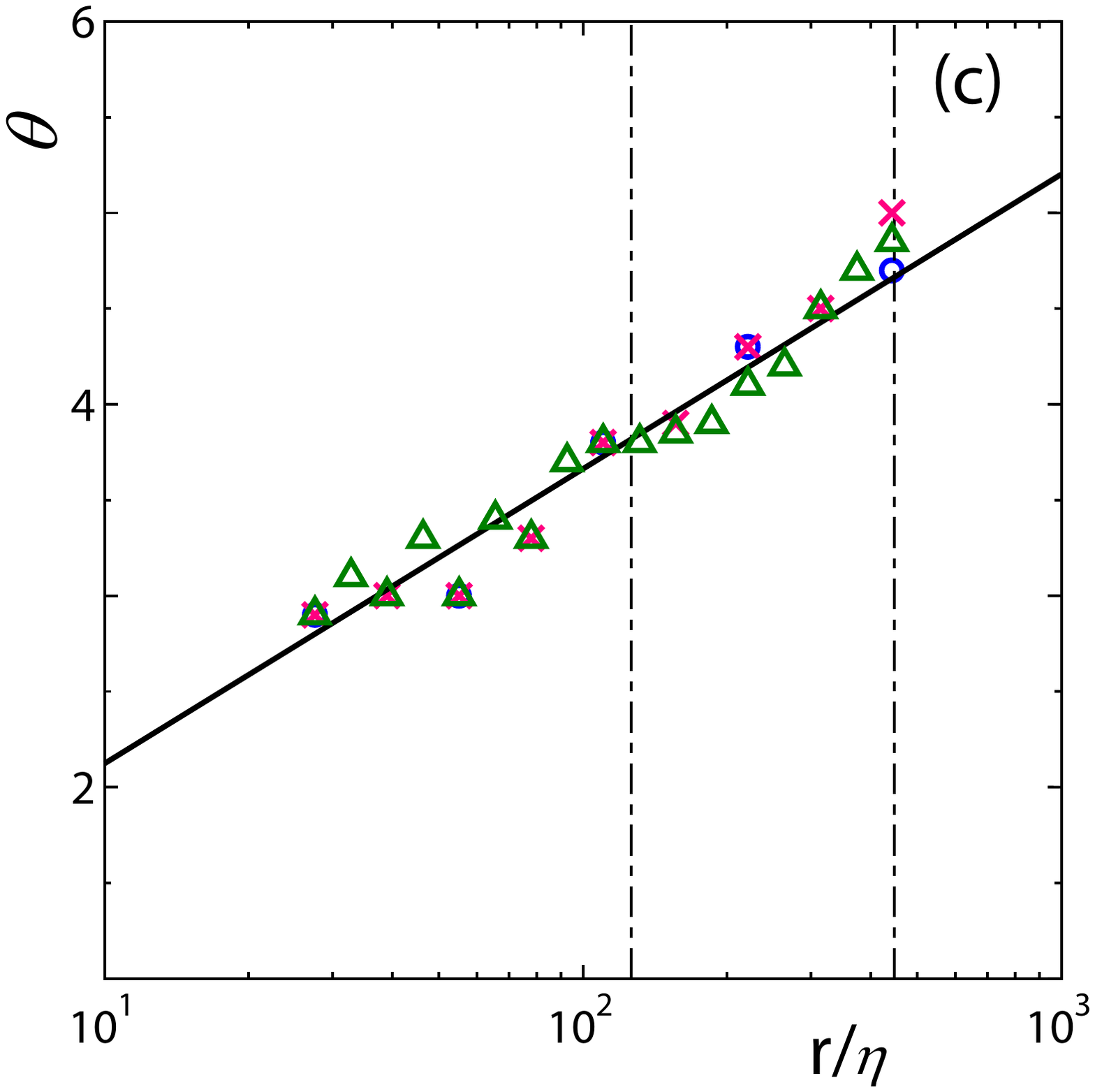}}%
\hspace*{0.5cm}
\resizebox*{7.0cm}{!}{\includegraphics{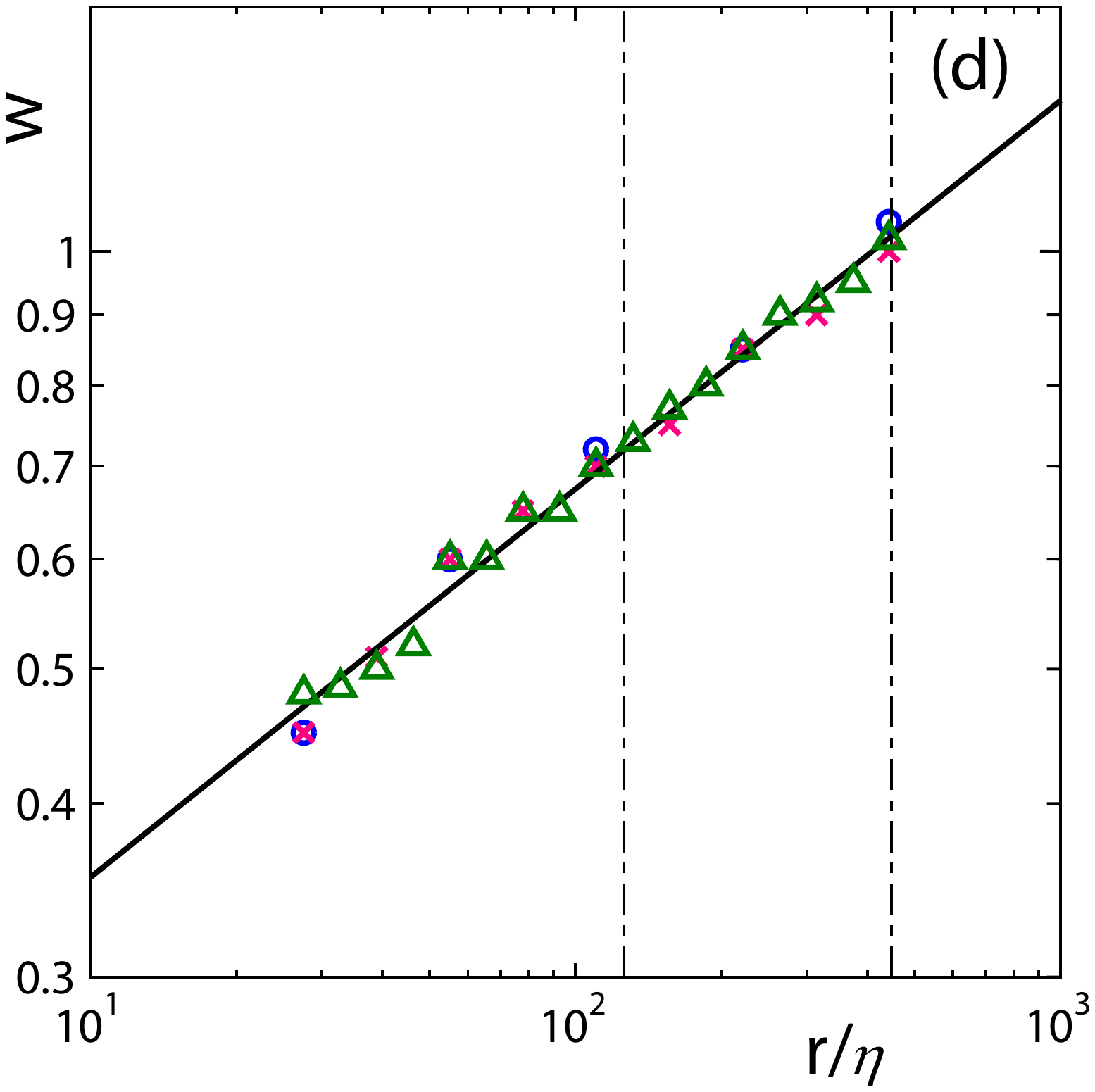}}%
\vspace{-2.5cm}
\caption{The $r/\eta$ ($=\ell_n/\eta$) dependence of the parameters 
(a) $\tilde{n}$, (b) $q^\prime$, (c) $\theta$ and (d) $w$ 
for the PDFs of energy dissipation rates, extracted from the whole of 
$4096^3$ DNS region.
In each figure, the results for $\delta=2^{1/4}$ ($\triangle$),
$\delta=2^{1/2}$ ($\times$) and $\delta=2$ ($\bigcirc$) are displayed altogether. 
The line in each figure is obtained by the method of least squares using
the data points for every $\delta$.
The obtained formula is 
(a) $\tilde{n}=-2.30\ \log_{10} (r/\eta)+8.37$,
(b) $q^\prime=3.71 \times 10^{-2}\ \log_{10}(r/\eta) + 1.00$,
(c) $\theta = 1.54\ \log_{10}(r/\eta) +0.585$ and
(d) $\log_{10} w = 0.280\ \log_{10} (r/\eta) + \log_{10} 0.186$.
The inertial range is the region between the vertical 
dash-dotted lines.
\label{fig: r-dependence of n, qp, theta, w}}
\end{center}
\end{figure}

\begin{figure}
\begin{center}
\vspace*{-2.2cm}
\includegraphics[height=.46\textheight]{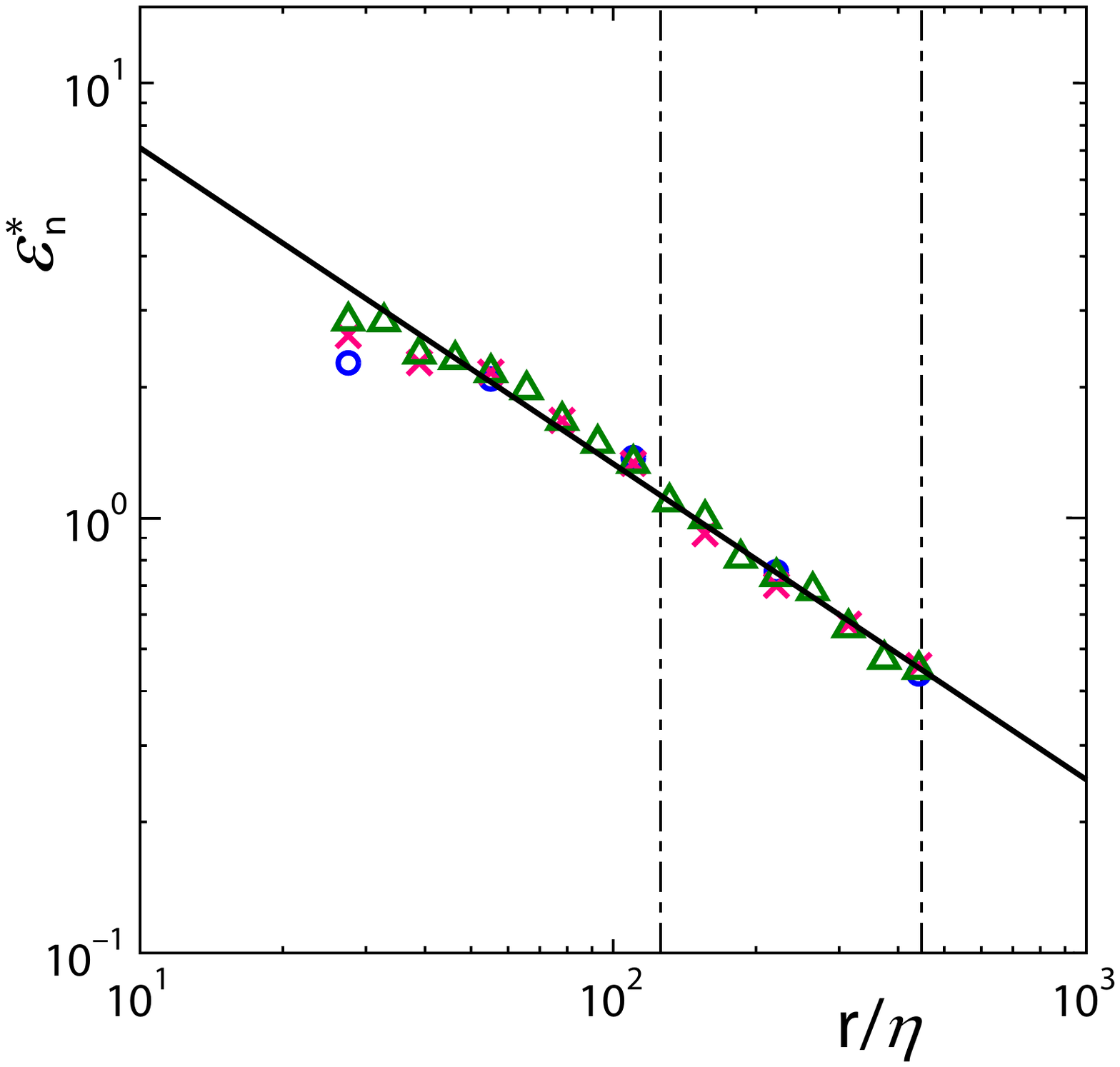}
\vspace{-2.5cm}
\caption{The $r/\eta$ ($=\ell_n/\eta$) dependence of the connection point 
$\varepsilon_n^*$ for PDFs of energy dissipation rates, extracted from the whole of 
$4096^3$ DNS region.
The results for $\delta=2^{1/4}$ ($\triangle$),
$\delta=2^{1/2}$ ($\times$) and $\delta=2$ ($\bigcirc$) are displayed altogether. 
The line 
$\log_{10} \varepsilon_{n}^{*} = -0.727\ \log_{10}(r/\eta) + \log_{10}37.8$
is obtained by the method of least squares using
the data points for every $\delta$.
The inertial range is the region between the vertical 
dash-dotted lines.
\label{fig: r-dependence of epsilon*}}
\end{center}
\end{figure}

Fig.~\ref{fig: r-dependence of n, qp, theta, w} and Fig.~\ref{fig: r-dependence of epsilon*}
give the dependence of 
$\tilde{n}$, $q^\prime$, $\theta$, $\ln w$ and $\ln \varepsilon_n$ on $r/\eta$ ($= \ell_n/\eta$)
for $\delta=2^{1/4}$ ($\triangle$), for $\delta=2^{1/2}$ ($\times$) 
and for $\delta=2$ ($\bigcirc$) extracted from the series of PDFs 
(see Fig.~\ref{fig:pdf edr delta_2^1/4,^1/2,^1}).
Here, $\tilde{n}$ is defined by $\tilde{n} = n \ln \delta$ with which $\ell_n$ 
introduced in (\ref{def of delta}) reduces to $\ell_n = \ell_0 \me^{-\tilde{n}}$.
The solid lines in Fig.~\ref{fig: r-dependence of n, qp, theta, w} (a), (b), (c), (d)
and in Fig.~\ref{fig: r-dependence of epsilon*} represent the empirical formulae 
given, respectively, by
\bea
\tilde{n} \aeq -0.998\ \ln(r/\eta)+8.37,
\label{extracted n tilde edr}
\\
q^\prime \aeq 1.61 \times 10^{-2}\ \ln (r/\eta)+1.00,
\label{extracted q prime edr}
\\
\theta \aeq 0.668\ \ln (r/\eta)+0.585,
\label{extracted theta edr}
\\
\ln w \aeq 0.280\ \ln (r/\eta) - 1.68,
\label{extracted w edr}
\\
\ln \varepsilon_{n}^{*} \aeq -0.727\ \ln(r/\eta) + 3.63,
\label{extracted epsilon edr}
\eea
which are obtained by the method of least squares using in each figure all the data points for 
$\delta = 2^{1/4}$, $2^{1/2}$ and $2$ altogether.
The results given in Fig.~\ref{fig: r-dependence of n, qp, theta, w}
and Fig.~\ref{fig: r-dependence of epsilon*} prove the correctness of the assumption that 
the fundamental quantities of turbulence are independent of $\delta$.
Note that, in the captions of these figures, we chose the base of logarithmic function 
in the empirical formulae to be 10 which corresponds to the axes of figures.

With the PDFs of high resolution, especially, for the center part 
given in Fig.~\ref{fig:pdf edr delta_2^1/4,^1/2,^1}, we succeeded to get 
the correct empirical formulae (\ref{extracted n tilde edr}), 
(\ref{extracted q prime edr}), (\ref{extracted theta edr}), (\ref{extracted w edr})
and (\ref{extracted epsilon edr}).
The independence of $\tilde{n}$ from $\delta$ ensures
the uniqueness of the PDF of $\alpha$ for any value of $\delta$
when the intermittency exponent $\mu$ has been settled.
There appears big difference in the parameters $q^\prime$, 
$\theta$ and $w$ responsible for the center-part PDFs compared with the results 
in \cite{AA18} obtained by PDFs with low resolution.
It turns out that $q^\prime \simeq 1.08$ but depending slightly on $r/\eta$ with 
positive slope for the high resolution (see Fig.~\ref{fig: r-dependence of n, qp, theta, w} (b)),
whereas it had negative slope for the low resolution (see Fig.~6 in \cite{AA18}).
The scattering of $\theta$ in the dependence of $r/\eta$ has been reduced 
(compare Fig.~\ref{fig: r-dependence of n, qp, theta, w} (c) with Fig.~7 in \cite{AA18}).
It is found that $\ln w$ depends lineally on $\ln (r/\eta)$ which is observed also 
in the analysis of PDFs for energy dissipation rates extracted from 
the turbulence in a wind tunnel \cite{AA19}, while $w$ was almost constant 
in the analyses of \cite{AA18} with less resolution.
The connection point $\alpha^*$ is adjusted in order for the best fit of the PDF 
around the region between the peak and the connection point. 
Note that the region $\alpha \leq \alpha^*$ ($\alpha > \alpha^*$) corresponds 
to the tail (center) part of the PDF.
It is revealed that the value $\alpha^*$ satisfies $\alpha^* \simeq 0.2$ for
all the data points with different values of $r/\eta$ 
(see Table~\ref{pdf edr parameters1}),
which proves the assumption that the center part $\Pi_{{\rm cr}}^{(n)}(\varepsilon_n)$
is constituted by two contributions, one from the coherent contribution
$\Pi_{{\rm S}}^{(n)}(\varepsilon_n)$ 
and the other from the incoherent contribution $\Delta \Pi^{(n)}(\varepsilon_n)$,
and that almost all the contribution to the tail part $\hat{\Pi}_{{\rm tl}}^{(n)}(\varepsilon_n)$
comes from the coherent motion of turbulence.
Remember that the energy dissipation rate becomes singular for $\alpha < 1$.

The comparison of the extracted formula (\ref{extracted n tilde edr}) for $\tilde{n}$
with the theoretical relation 
$
\tilde{n} = -\ln (r/\eta) +\ln (\ell_0/\eta)
$
provides us with the estimation $\ell_0/\eta = 4.31 \times 10^3$.
Since the smallest grid spacing is $3\eta$~\cite{Kaneda-Ishihara05}, 
$\ell_0/3\eta = 1.44 \times 10^3$ provides us with the number of grids 
corresponding to $\ell_0$.
Note that $\ell_0/\eta$ is about 2 times larger than the integral length 
$L/\eta$ of the system~\cite{Kaneda-Ishihara05}.
It should be noted here that we observed $\ell_0/\eta \simeq L/\eta$ for the case of 
experimental turbulence in a wind tunnel~\cite{AA19}.

\section{Analysis of PDFs taken from partial DNS regions
\label{dense and rare PDFs}}

We are analyzing in this section the PDFs of energy dissipation rates created
from the snapshot data in partial regions of the size $512^3$ which are obtained 
by cutting the whole of $4096^3$ DNS region into $512$ pieces.
We will refer to the PDF created from the partial region as p-PDF in the following.
Among $512$ partial DNS regions, we select in this paper two DNS regions, i.e.,
one has a maximum enstrophy and the other a minimum enstrophy, 
and study the series of p-PDFs obtained in these regions.
We call the p-PDF created from the partial DNS region with maximum (minimum) enstrophy 
max-PDF (min-PDF).

\begin{figure}
\begin{center}
\vspace*{-2.3cm}
\resizebox*{7.5cm}{!}{\includegraphics{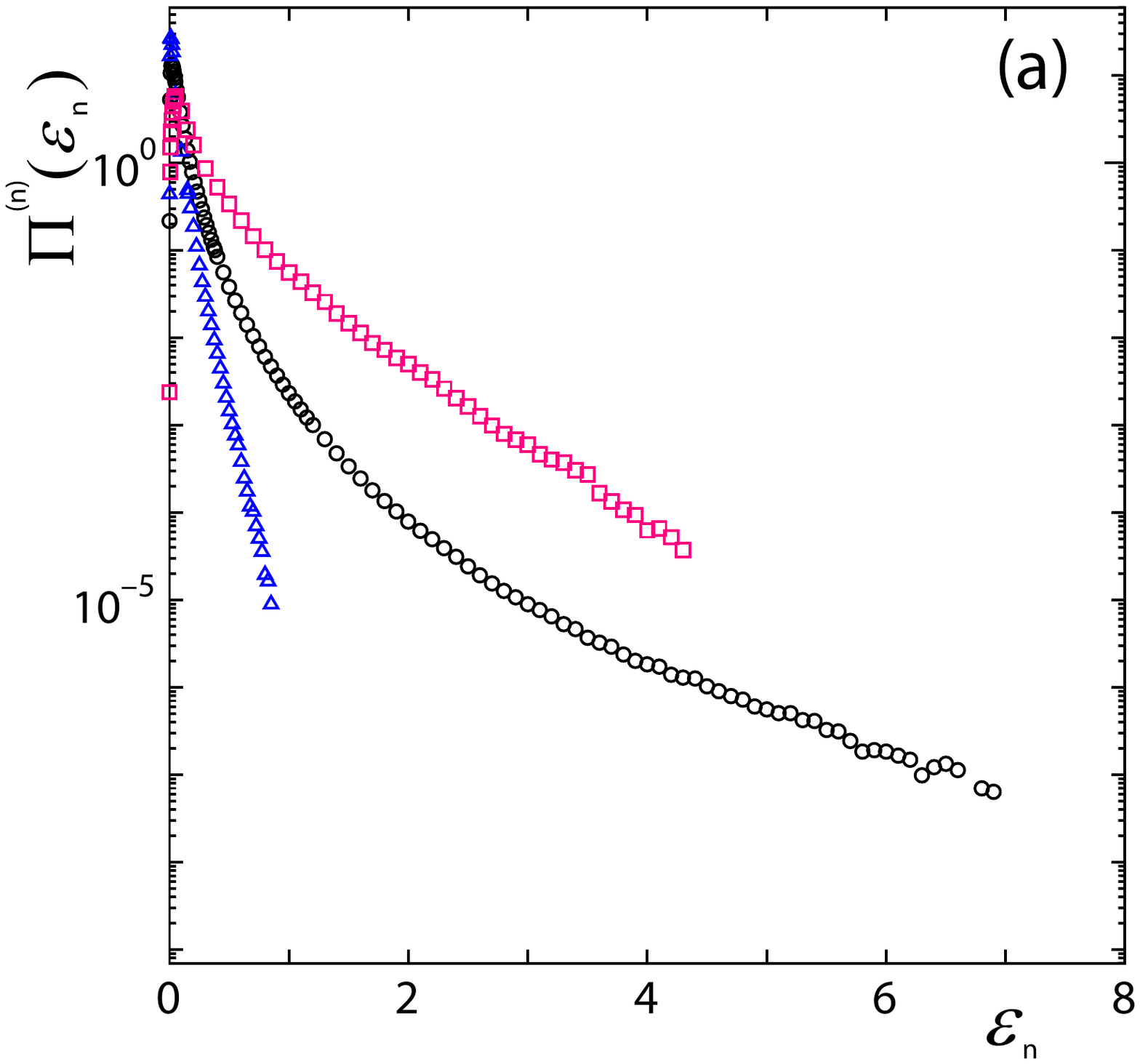}}%
\hspace*{-1.5cm}
\resizebox*{7.5cm}{!}{\includegraphics{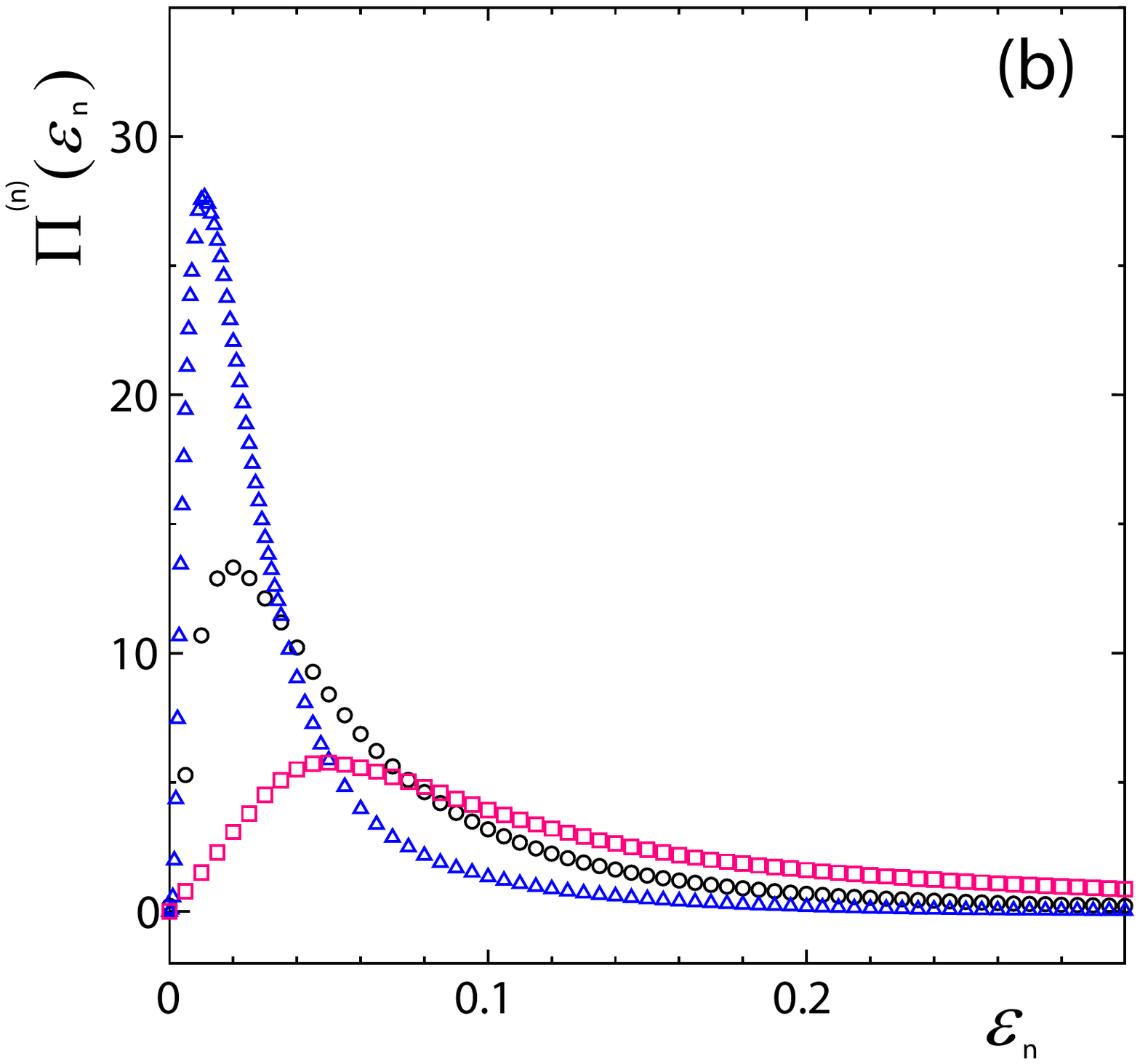}}%
\vspace{-2.6cm}
\caption{Comparison of the max-PDF ($\square$), the min-PDF ($\triangle$) and
the w-PDF ($\bigcirc$) for $r/\eta = 55.1$.
The horizontal axes represent the energy dissipation rates $\varepsilon_n$, 
and the vertical axes do the PDFs on (a) log and (b) linear scale.
\label{fig:minnax-pdf for a r/eta}}
\end{center}
\end{figure}

The max-PDF ($\square$) and the min-PDF ($\triangle$) for the case of $r/\eta = 55.1$
are displayed as functions of $\varepsilon_n$ 
in Fig.~\ref{fig:minnax-pdf for a r/eta} with the vertical axes in
(a) log scale and (b) linear scale.
In the figure, we put the PDF ($\bigcirc$) created from the whole of $4096^3$ DNS region, 
which we call w-PDF in the following, for the same value of $r/\eta$ as a reference.
We observe by comparing with the w-PDF that the proportion of the probability density 
for the max-PDF (the min-PDF) is shifted to the tail-part PDF (the center-par PDF). 
It is reasonable in the sense that since the vortexes are distributed dense (sparse) 
in the partial region with maximum (minimum) enstrophy, the proportion of 
the intermittent coherent (the fluctuating incoherent) fluid motion should be large.

\begin{figure}
\begin{center}
\vspace*{-0.65cm}
\resizebox*{7.5cm}{!}{\includegraphics{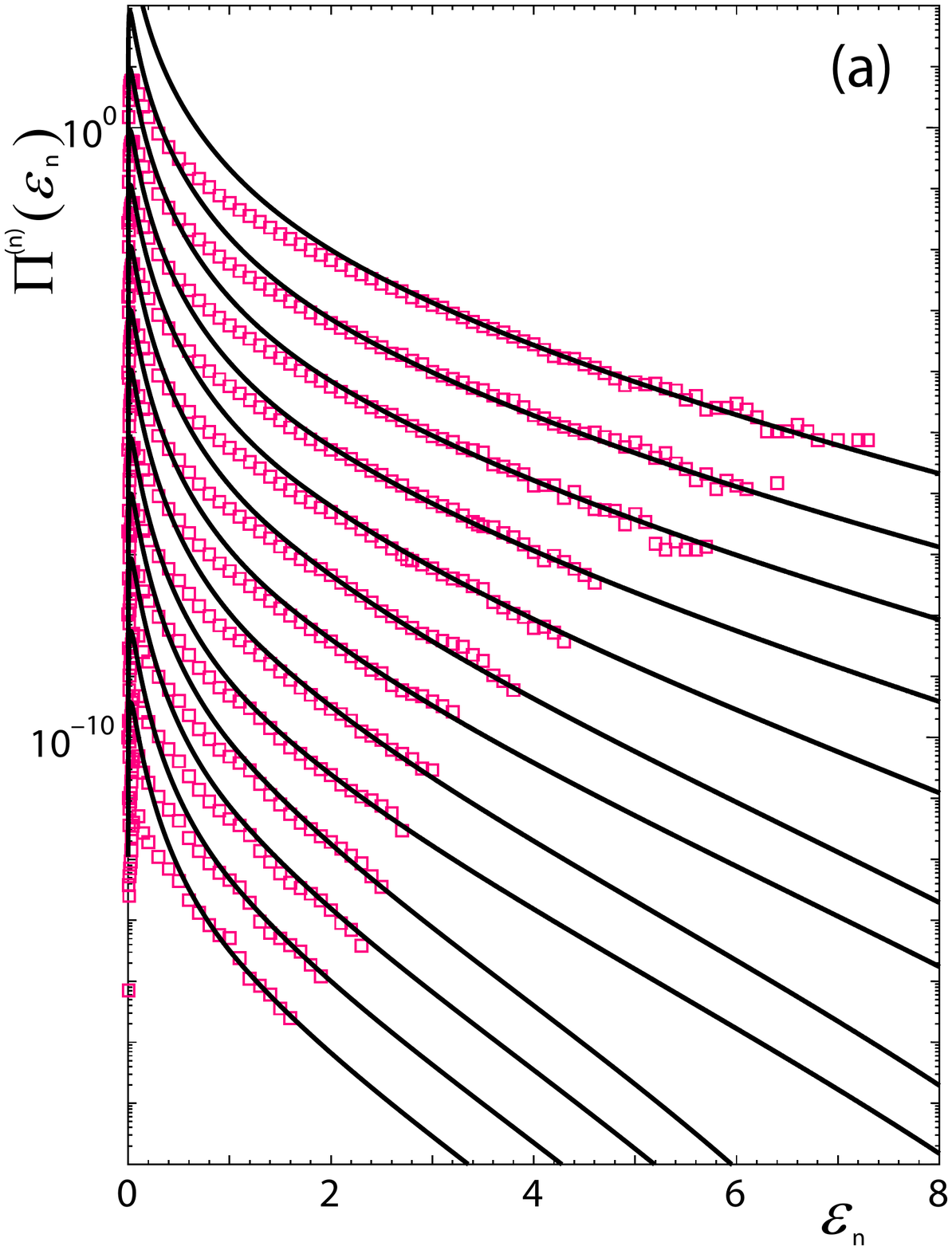}}%
\resizebox*{7.5cm}{!}{\includegraphics{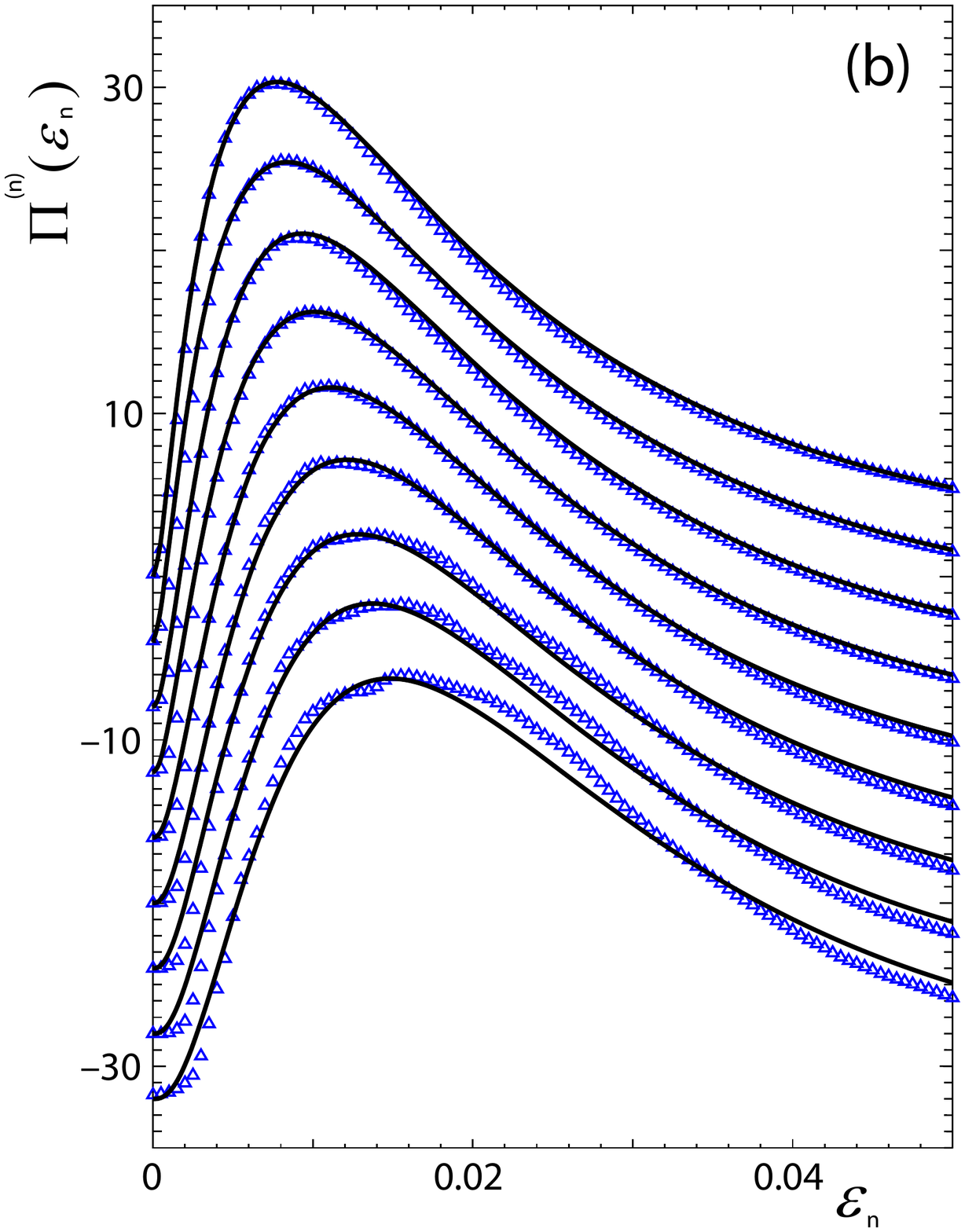}}%
\vspace{-2.0cm}
\caption{Analyses of (a) the max-PDFs ($\square$) and (b) the min-PDFs ($\triangle$) 
of energy dissipation rates for $\delta = 2^{1/4}$.
For better visibility, the PDFs in each figure are shifted along the vertical axes 
by the amount $-1$ in (a), $-4$ in (b) according to the values 
$r/\eta = 27.5$, $32.8$, $38.9$, $46.3$, $55.1$, $65.6$, $78.0$, $92.7$, $110$, $131$, $156$,
$186$ and $221$ for (a), and
$r/\eta = 27.5$, $32.8$, $38.9$, $46.3$, $55.1$, $65.6$, $78.0$, $92.7$ and $110$ for (b), 
successively, from the top to the bottom.
Solid lines are the w-PDFs drawn in Fig.~\ref{fig: r-dependence of n, qp, theta, w} 
after appropriate cooking procedures, i.e., (a) multiplying some factor $\chi$ to w-PDF,
(b) scaling w-PDF as $\breve{\Pi}^{(n)}(\breve{\varepsilon}_n)$ in (\ref{breve Pi}),
whose recipe is given in Fig.~\ref{fig:recipe} (see the main text for detail).
\label{fig:minnax-pdf}}
\end{center}
\end{figure}

\begin{figure}
\begin{center}
\vspace*{-1.8cm}
\resizebox*{15.0cm}{!}{\includegraphics{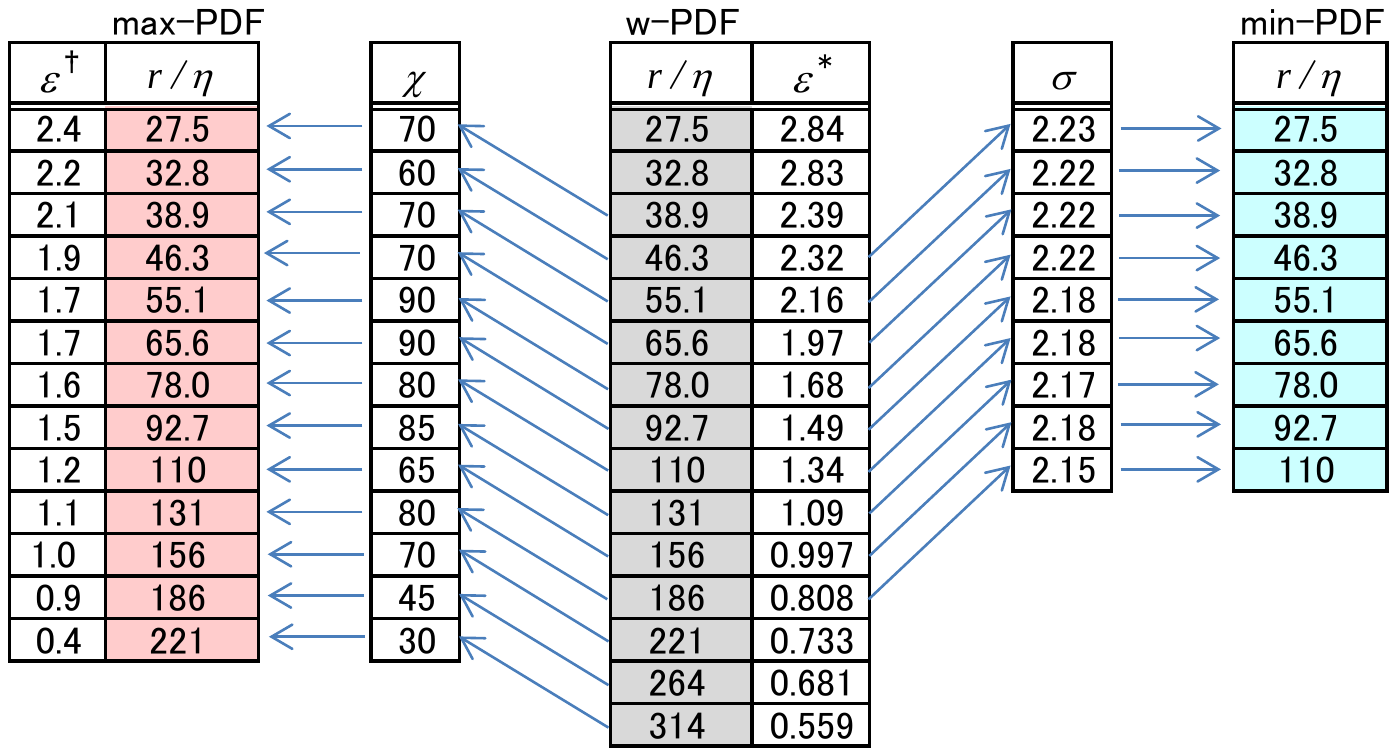}}%
\vspace{-13.5cm}
\caption{Recipe to analyze the max-PDFs and the min-PDFs by means of the w-PDFs.
The arrows indicate that which w-PDF with a specific $r/\eta$ at the origin of an arrow
can adjust the max-PDF or the min PDF at the point of the arrow, respectively, 
by the magnification factor $\chi$ and by the scaling factor $\sigma$.
$\varepsilon^\dagger$ is the point from which a max-PDF ($\square$) and a solid line
representing a w-PDF in Fig.~\ref{fig:minnax-pdf} (a) multiplied by the factor $\chi$ 
start to merge. 
The values of $\varepsilon^*$ associated with w-PDF for $\delta = 2^{1/4}$ are listed 
here again for convenience.
\label{fig:recipe}}
\end{center}
\end{figure}

The max-PDFs ($\square$) and the min-PDFs ($\triangle$) of energy dissipation rates 
for $\delta = 2^{1/4}$ are given, respectively, in Fig.~\ref{fig:minnax-pdf} (a) and (b).
For better visibility, the PDFs in each figure are shifted along the vertical axes 
by the amount $-1$ in (a), $-4$ in (b) according to the values 
$r/\eta = 27.5$, $32.8$, $38.9$, $46.3$, $55.1$, $65.6$, $78.0$, $92.7$, $110$, $131$, $156$,
$186$ and $221$ for (a), and
$r/\eta = 27.5$, $32.8$, $38.9$, $46.3$, $55.1$, $65.6$, $78.0$, $92.7$ and $110$ for (b), 
successively, from the top to the bottom.
Solid lines are the w-PDFs drawn in Fig.~\ref{fig: r-dependence of n, qp, theta, w}
(a), (b) after appropriate cooking procedures whose recipe is listed in Fig.~\ref{fig:recipe}.
From Fig.~\ref{fig:minnax-pdf} (a), it was revealed that the tail-part of max-PDFs 
can be adjusted with the slope of the tail-part of w-PDF with a specific value of 
$r/\eta$ magnified by some factor $\chi$, 
and that the value of $\varepsilon_n^\dagger$ at which a max-PDF ($\square$) and
a solid line representing a w-PDF in Fig.~\ref{fig:minnax-pdf} (a) start to overlap 
is quite close to the connection point $\varepsilon_n^*$ associated to the w-PDF 
(see Fig.~\ref{fig:recipe}).
From Fig.~\ref{fig:minnax-pdf} (b), it was found that the center-part of min-PDFs 
can be adjusted by the scaled PDF, $\breve{\Pi}^{(n)}(\breve{\varepsilon}_n)$,
which is introduced through
\be
\breve{\Pi}^{(n)}(\breve{\varepsilon}_n)\ d \breve{\varepsilon}_n
= \Pi^{(n)}(\varepsilon_n)\ d \varepsilon_n, 
\qquad \breve{\varepsilon}_n = \varepsilon_n / \sigma
\label{breve Pi}
\ee
with an appropriate scaling factor $\sigma$ (see Fig.~\ref{fig:recipe}).

In Fig.~\ref{fig:recipe}, we listed the values of the magnification $\chi$ 
which is necessary to analyze the max-PDFs in Fig.~\ref{fig:minnax-pdf} (a), 
and of the scaling factor $\sigma$ which is necessary for the scaled PDF 
$\breve{\Pi}^{(n)}(\breve{\varepsilon}_n)$ to analyze the min-PDFs 
in Fig.~\ref{fig:minnax-pdf} (b).
The arrows indicate that which w-PDF with a specific value $r/\eta$ 
at the origin of an arrow can adjust the max-PDF or the min-PDF 
at the point of the arrow.
In between the arrows, there inserted the magnification factor $\chi$ for the case of 
max-PDF or the scale factor $\sigma$ for the case of min-PDF (see also Fig.~\ref{fig:minnax-pdf}).

From Fig.~\ref{fig:recipe}, we notice at least two remarkable outcomes.
One is the fact that the value of $\varepsilon_n^\dagger$ associated with max-PDF is almost
equal to the values of the connection point $\varepsilon_n^*$, which may indicates that
the division of the tail part and the center part of PDF within A\&A model is reasonable.
The other is the fact that the scaling factor $\sigma$ has almost a common value around 
$2.19$, which may indicate that there exists a beautiful statistics associated with 
the incoherent fluctuation part of turbulent motion.
The latter may be related to the beautiful scaling behaviors given by
the empirical formulae (\ref{extracted q prime edr}), (\ref{extracted theta edr}) 
and (\ref{extracted w edr}) for the parameters associated with the center-part PDFs.

\section{Conclusion
\label{conclusion}}

We investigated quite accurately the PDFs of energy dissipation rates created 
in a high resolution by cooking the snapshot data taken from the whole of $4096^3$ DNS region
by the theoretical formula for PDF derived within A\&A model of MPDFT whose contents
are compactly given in the present paper. 
Analyzing the obtained high-resolution w-PDFs, we derived the empirical formulae of 
the parameters consisting, especially, the central part of the theoretical PDF 
in their most precise forms for the first time.
It was revealed that $\tilde{n}$, $q'$, $\theta$, $\ln w$ and $\ln \varepsilon_n^*$ 
(therefore, $\alpha^*$) are independent of $\delta$ thanks to the new scaling relation 
(\ref{new scaling relation}), and that they show scaling behaviors extending to the regions 
with smaller $r/\eta$ values from the inertial range.
By making use of the w-PDFs, we also succeeded to extract the some attractive informations
contained in the max-PDFs and the min-PDFs (see Fig~\ref{fig:recipe}): 
1) We can find a w-PDF whose tail part can adjust the slope of the tail-part of a max-PDF
with appropriate magnification factor $\chi$.
2) The value of $\varepsilon_n^\dagger$ at which the w-PDF multiplied by $\chi$ starts
to overlap the tail part of the max-PDF coincides well with the connection point
$\varepsilon_n^*$ for the theoretical w-PDF.
3) The center part of the min-PDFs can be adjusted, quite accurately, by the scaled PDFs 
defined by (\ref{breve Pi}) with a scale factor $\sigma$.
It is attractive that the value of $\sigma$ is almost common to every min-PDFs 
with different values $r/\eta$.

There is no theoretical prediction yet, which is based on an ensemble theoretical aspect 
or on a dynamical aspect starting with the N-S equation, to produce 
the formula for the center part PDF that represents the contributions both 
of the coherent turbulent motion providing 
intermittency and of incoherent fluctuations (background flow) around 
the coherent motion.
The discoveries given above may provide us with a correct pathway to formulate 
a dynamical theory which produces, properly, the formula for the center part of PDFs 
starting with the N-S equation.
The discoveries open a new door to separate the two elements of turbulence, i.e.,
the coherent motion and the incoherent motion, and may lead us to a appropriate new method 
to make each element visible, separately, in the near future.
A study to this direction is now in progress,
and will be reported elsewhere in the near future.

Let us close this paper by noting the following comments.
It was found that the new scaling relation (\ref{new scaling relation}) is deeply 
related to the $\delta$-scale Cantor sets~\cite{Halsey-Jensen-Kadanoff-Procaccia-Shraiman86} 
created from $\delta^\infty$ periodic orbits~\cite{Motoike-TA11}.
In this respect, it may be reasonable to interpret that $n$ introduced in (\ref{def of delta})
represents the number of stages in the $\delta$-scale Cantor sets
since $n$ increases approximately by one for every $\delta$ 
(see Table~\ref{pdf edr parameters1}).
On the other hand, we observe that the $\tilde{n}$ is independent of $\delta$, and 
therefore it may be appropriate to interpret that $\tilde{n}$ is a good number
representing a number of steps associate with the energy cascade model.

\section*{Acknowledgment}

The authors (T.A. and N.A.) would like to thank Prof.~T.~Motoike, 
Dr.~K.~Yoshida and Mr.~M.~Komatsuzaki for fruitful discussions.
They are also grateful to Dr.~H.~Mouri for useful comments.

\setlength{\baselineskip}{5pt}

\end{document}